\documentclass[aps,prc,preprint,tightenlines,superscriptaddress,showpacs,%
amssymb,byrevtex,nofootinbib]{revtex4}
\usepackage[dvips]{graphicx,color}
\usepackage{amsmath}
\usepackage{mathrsfs}
\usepackage{dcolumn}
\usepackage{epsfig}
\usepackage{bm}

\newcommand{\non}{\nonumber}
\newcommand{\fm}{\,{\rm fm}}
\newcommand{\mev}{\,{\rm MeV}}

\newcommand{\ZZ}{\mathscr{Z}}

\newcommand{\KKbar}{{K\bar{K}}}
\newcommand{\pipi}{{\pi\pi}}

\begin{document}
\title{Finite-volume Hamiltonian method for coupled channel interactions in lattice QCD}
\author{Jia-Jun Wu}
\affiliation{Physics Division, Argonne National Laboratory, Argonne, Illinois 60439, USA}
\author{T.-S. H. Lee}
\affiliation{Physics Division, Argonne National Laboratory, Argonne, Illinois 60439, USA}
\author{A. W. Thomas}
\affiliation{Special Research Center for the Subatomic Structure of Matter (CSSM),
School of Chemistry and Physics,  University
of Adelaide Adelaide 5005, Australia}
\affiliation{ARC Center of Excellence for Particle Physics at Terascale,School of Chemistry and Physics,  University
of Adelaide Adelaide 5005, Australia}
\author{R. D. Young}
\affiliation{Special Research Center for the Subatomic Structure of Matter (CSSM),
School of Chemistry and Physics,  University
of Adelaide Adelaide 5005, Australia}
\affiliation{ARC Center of Excellence for Particle Physics at Terascale,School of Chemistry and Physics,  University
of Adelaide Adelaide 5005, Australia}
\begin{abstract}
  Within a multi-channel formulation of $\pi\pi$ scattering, we
  investigate the use of the finite-volume Hamiltonian approach to
  resolve scattering observables from lattice QCD spectra.
  The asymptotic matching of the well-known L\"uscher
    formalism encodes a unique finite-volume spectrum. Nevertheless,
    in many practical situations, such as coupled-channel systems, it
    is advantageous to interpolate isolated lattice spectra in order
    to extract physical scattering parameters. Here we study the
    use of the Hamiltonian framework as a parameterisation that can be
    fit directly to lattice spectra.
  We find that with a modest amount of lattice data,
  the scattering parameters can be reproduced rather well, with only a
  minor degree of model dependence.
\end{abstract}

\maketitle

\section{Introduction}
Lattice QCD studies are making tremendous progress in resolving the
excitation spectrum of QCD
\cite{Dudek:2011tt,Edwards:2011jj,Menadue:2011pd,Mahbub:2012ri,Dudek:2012xn}. By
the nature of the finite-volume and Euclidean time aspects of the
lattice formulation, it is impossible to directly simulate scattering
processes. The established way to extract of scattering information
from lattice simulations is the L\"uscher method
\cite{Luscher:1986pf,Luscher:1990ux}.  For the case of elastic 2-body
scattering, L\"uscher identified that the
%
finite volume eigenstates are uniquely determined in terms of the on-shell scattering parameters
%
(up to exponentially suppressed corrections associated
with quantum fluctuations of the lightest degrees of freedom in the
system).
%
  While the spectrum is determined uniquely, there are technical
  challenges associated with inverting a given lattice spectrum to
  determine scattering observables. One of these issues arises from
  the fact that the full rotational group is broken down by the
  geometry of the lattice boundary conditions. As a consequence, partial wave
  mixing is unavoidable in lattice simulations and eigenstates on the
  finite volume do not correspond to definite eigenstates of the
  continuum rotation group. There has been significant work in
  previous years addressing this issue,
  eg.~Refs.~\cite{Doring:2011nd,Dudek:2012gj,Doring:2012eu,Dudek:2012xn,Briceno:2013bda}.

In the present work, we focus our attention of the study of inelastic
scattering channels.
  The
generalisation of the L\"uscher
formalism to incorporate inelastic channels was developed by He, Feng
and Liu \cite{He:2005ey}, and
continues to be
the topic of
considerable further investigations
and extensions
\cite{Lage:2009zv,Bernard:2010fp,MartinezTorres:2011pr,Doring:2011nd,Hansen:2012tf,Briceno:2012yi,Doring:2012eu,Li:2012bi,Guo:2012hv,Briceno:2013bda}.
In addition to the issue of partial wave mixing, coupled-channel systems are further complicated by the multi-component nature of the $S$-matrix.
%
For example,
neglecting the angular momentum mixing,
for the
case of two coupled channels on a given volume,
a single energy
  eigenstate is related to
three asymptotic scattering parameters
(i.e. two phase shifts and an inelasticity).
Therefore the only way to uniquely identify all three parameters would
be to search for
near-
coincident energy eigenstates at either
different volumes or with different momentum boosts of the system
\cite{Guo:2012hv}.
In practice, such a ``pointwise'' extraction is
only anticipated to have limited applicability.
%
Alternatively, one requires some form of
interpolation which can reproduce the scattering parameters with a
limited set of lattice simulation results. In the present work, we
extend a recently developed finite-volume Hamiltonian formalism
\cite{Hall:2013qba} to a coupled-channel system.
%
%
The necessary equivalence with the L\"uscher formalism is
  numerically established.
%
Further, we investigate the inversion problem of extracting
the phase shifts and inelasticity from a finite set of pseudo lattice
data. We find that all three scattering parameters can be reliably
reproduced by directly constraining the parameters of the model to the
finite volume spectra. In the energy region constrained by the fits,
the extracted phase shifts and inelasticity show only a mild
sensitivity to the precise form of the model.

To facilitate the exploration of LQCD spectra, our analysis is based
upon a two-channel Hamiltonian formulation which is constructed by
fitting the available $\pi\pi$ scattering phase shifts data in the
$J^{IP}=0^{0+}, 1^{1-}$ partial waves. The explicit channels included
are $\pi\pi$ and the inelasticity associated with $K\bar{K}$
production.
%
With the present manuscript being focussed primarily on the
  influence of the inelastic channel, we do not consider the issues
  associated with angular momentum mixing.
%

In section II, we write down a multi-channel formulation for
constructing several model Hamiltonians from fitting the $\pi\pi$ scattering data.
The model with only the $\pi\pi$ channel is used in section III to recall the finite-box
Hamiltonian method developed in Ref.~\cite{Hall:2013qba} and  to
examine the correspondence with L\"uscher's formula. In section IV, we use the
 model with $\pi\pi$ and $K\bar{K}$ channels to show that the finite-box Hamiltonian
approach is equivalent to the approach based on the two-channel L\"uscher's method developed in
Ref.~\cite{He:2005ey}. In section IV, we compare the LQCD efforts needed to apply
 the finite-box Hamiltonian approach and
the approach based on L\"uscher's method.
Our predictions of the spectra for testing LQCD results for $\pi\pi$ scattering in
the $J^{IP}=0^{0+}, 1^{1-}$ partial waves are presented in section V. In section VI, we give
a summary and discuss possible future developments.

\section{Model Hamiltonian for $\pi\pi$ scattering}

The Hamiltonian with only vertex interactions, such as $\Delta
\leftrightarrow \pi N$ considered in Ref.\cite{Hall:2013qba}, is the
simplest example within the general multi-channel formulation,
inspired by the cloudy bag model ~\cite{Theberge:1980ye,Thomas:1982kv} and
developed in Ref.~\cite{Matsuyama:2006rp} for investigating the
nucleon resonances \cite{Kamano:2013iva} and meson resonances
\cite{Kamano:2011ih}.  For investigating the finite-box Hamiltonian
approach in this work, it is useful to recall the formulation of
Refs.~\cite{Matsuyama:2006rp,Kamano:2011ih} in order to write down a general
Hamiltonian for $\pi\pi$ scattering.

Following Refs.~\cite{Matsuyama:2006rp,Kamano:2011ih}, we assume that
 $\pi\pi$ scattering can be described by vertex interactions and
two-body potentials. In the rest frame, the model Hamiltonian of a
meson-meson system takes the following {\em energy-independent} form
\begin{eqnarray}
H = H_0 + H_I.
\label{eq:h}
\end{eqnarray}
The non-interacting part is
\begin{eqnarray}
H_0 =\sum_{i=1,n} |\sigma_i\rangle m^{0}_i \langle\sigma_i|
+ \sum_{\alpha} \int d\vec{k} |\alpha(\vec{k})\rangle[\sqrt{m_{\alpha_1}^2 + \vec{k}^{\,\,2}_{\alpha_1}} +
\sqrt{m_{\alpha_2}^2 + \vec{k}^{\,\,2}_{\alpha_2}}] \langle\alpha(\vec{k})|,
\label{eq:h0}
\end{eqnarray}
where $\sigma_i$ is the $i$-th bare particle with mass $ m^{0}_i$,
$\alpha = \pi\pi, K\bar{K}, \pi\eta, \cdot\cdot$ denotes the channels
included, and $m_{\alpha_i}$ and $\vec{k}_{\alpha_i}$ are the mass and
the momentum of the $i$-th particle in the channel $\alpha$,
respectively. In the considered center of mass system, we obviously
have defined $\vec{k}_{\alpha_1}=-\vec{k}_{\alpha_2}=\vec{k}$.

The interaction Hamiltonian is
\begin{eqnarray}
H_I = g + v,\label{eq:hi}
\end{eqnarray}
where $g$ is a vertex interaction describing the decays of the bare particles
into two-particle channels $\alpha, \beta, \ldots$
\begin{eqnarray}
g = \sum_{\alpha}\int d\vec{k} \sum_{i=1,n} \{  |\alpha(\vec{k})\rangle g^\dagger_{i,\alpha}(k) \langle i| +
|i\rangle g_{i,\alpha}(k) \langle \alpha(\vec{k})|\}
\label{eq:int-g}
\end{eqnarray}
and the direct two-particle-two-particle interaction is defined by
\begin{eqnarray}
v = \sum_{\alpha,\beta} \int d\vec{k} d\vec{k}'\, |\alpha(\vec{k})\rangle  v_{\alpha,\beta}(k,k') \langle \beta(\vec{k}')|.
\label{eq:int-v}
\end{eqnarray}

In each partial wave, the two particle scattering is then defined
by the following coupled-channel equations
\begin{align}
t_{\alpha,\beta}(k,k'; E) &= V_{\alpha,\beta}(k,k')\non\\
&\quad
+ \sum_{\gamma}\int _0^{\infty} k^{''\,\,2}dk^{''}
V_{\alpha,\gamma}(k,k'')\frac{1}{E-E_{\gamma_1}(k^{''}) - E_{\gamma_2}(k^{''})
+i\epsilon} t_{\gamma,\beta}(k^{''},k';E)
\label{eq:lseq-1}
\end{align}
where $E_{\gamma i}=\sqrt{k^{''\,\,2} + m^2_{\gamma_i}}$, and the coupled-channel
potentials are
\begin{eqnarray}
V_{\alpha,\beta}(k,k') = \sum_{i=1,n}g^*_{i,\alpha}(k)\frac{1}{E-m_i^0} g_{i,\alpha}(k')
+v_{\alpha,\beta}(k,k')
\label{eq:lseq-2}
\end{eqnarray}
with
\begin{eqnarray}
g_{i,\alpha}(k)&=&\langle i|g|\alpha(\vec{k})\rangle  \\
v_{\alpha,\beta}(k,k')&=&\langle \alpha(\vec{k})|v|\beta(\vec{k}')\rangle
\end{eqnarray}
We choose the normalization,
$\langle \alpha(\vec{k})|\beta(\vec{k}^{\,\,'})\rangle  = \delta_{\alpha,\beta}\delta (\vec{k}-\vec{k}^{\,\,'})$
such that the S-matrix in each partial-wave is
related to the T-matrix by
\begin{eqnarray}
 S_{\alpha,\beta}(E) = 1 +2 i
T_{\alpha,\beta}(k_{0\alpha},k_{0\beta};E)\label{eq:ST1}
\end{eqnarray}
with
\begin{eqnarray}
T_{\alpha,\beta}(k_{0\alpha},k_{0\beta};E)
=-\rho^{1/2}_{\alpha}(k_{0\alpha})
t_{\alpha,\beta}(k_{0\alpha},k_{0\beta};E) \rho^{1/2}_{\beta}(k_{0\beta})\label{eq:ST2}
\end{eqnarray}
where $k_{0\alpha}$ is the on-shell momentum for the channel $\alpha$
and the density of states is
\begin{eqnarray}
\rho_{\alpha}(k_{0\alpha})=\pi \frac{k_{0\alpha}E_{\alpha 1}(k_{0\alpha})
E_{\alpha 2}(k_{0\alpha})}{E_{\alpha 1}(k_{0\alpha})+E_{\alpha 2}(k_{0\alpha})}\label{eq:ST3}
\end{eqnarray}

In the following sections, we construct
(1) one-bare state and one-channel ($1b-1c$) models,
(2) one-bare state and two-channels ($1b-2c$) models,
and also
(3) two-bare states and two-channels ($2b-2c$) models.

%
\section{one bare state and one-channel}

In this section, we  consider a model which
has one bare state ($\sigma$) and one-channel ($\pi\pi$) to describe the isoscalar
$s$-wave $\pi\pi$ scattering
phase shifts up to the energy below the $K\bar{K}$ threshold.
The formulae for constructing this model, called $1b-1c$ model,
 can be obtained from  taking $n=1$ and $\alpha=\beta=\gamma=\pi\pi$
in section II.

\subsection{Model parameters}
\begin{figure}[htbp] \vspace{-0.cm}
\begin{center}
\includegraphics[width=0.49\columnwidth]{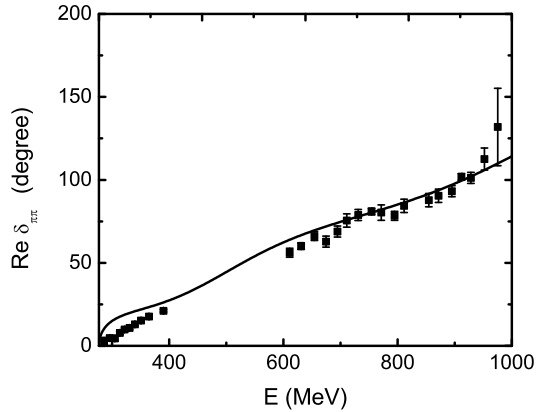}
\caption{The phase shifts of $\pi\pi$ scattering from the Model 1b-1c (cf. Table \ref{tab:expfit})
are compared with the data.}
\label{fg:expfit1}
\end{center}
\end{figure}

For simplicity, we parametrize the matrix elements of the interactions
in Eqs.(4) and (5) as
\begin{eqnarray}
\langle \sigma|g| \pi\pi(\vec{k}) \rangle &=&g_{\sigma, \pi\pi}(k) \nonumber \\
                      &=&\frac{g_{\pi\pi}}{\sqrt{\pi}}\frac{1}{(1+(c_{\pi\pi}\times k)^2)},\label{eq:gepp1}\\
\langle \pi\pi(\vec{k})|v|\pi\pi(\vec{k}')\rangle  &=&v_{\pi\pi, \pi\pi}(k, k')\nonumber \\
                       &=&\frac{G_{\pi\pi,\;\pi\pi}}{m^2_\pi}\times
\frac{1}{(1+(d_{\pi\pi}\times k)^2)^2}\times\frac{1}{(1+(d_{\pi\pi}\times k')^2)^2},\label{eq:vpp1}
\end{eqnarray}
where $k$ and $k'$ are the three momenta of $\pi$ in the center of
mass system.  By fitting the $\pi\pi $ phase shifts, the parameters,
$m_\sigma$, $g_{\pi\pi}$, $c_{\pi\pi}$, $G_{\pi\pi,\;\pi\pi}$ and
$d_{\pi\pi}$, of the model can be determined and are listed in the
column ``1b-1c'' in Table \ref{tab:expfit}. The calculated phase
shifts are compared with the data in Fig.~\ref{fg:expfit1}.
The model gives a reasonable description of the data and is sufficient for
exploring the systematics of the finite-volume Hamiltonian method.

\begin{table}[ht]
     \setlength{\tabcolsep}{0.15cm}
\begin{center}
\caption{The  parameters for the 1 bare state and 1 channel ($1b-1c$) model,
and the 1 bare state and 2 channel ($1b-2c$) model}
\begin{tabular}{ccccc}\hline
                           &1b-1c  &  & 1b-2c    &  \\
\hline
 $m_\sigma$(MeV)            &700.      &  & $700.00$   & \\
 $g_{\sigma\pi\pi}$         &1.6380    &  & $2.0000$   & \\
 $c_{\sigma\pi\pi}$(fm)     &1.0200    &  & $0.6722$   & \\
 $G_{\pi\pi,\;\pi\pi}$      &0.5560    &  & $2.4998$   & \\
 $d_{\pi\pi}$(fm)           &0.5140    &  & $0.2440$   & \\
 $g_{\sigma K\bar{K}}$      &  -       &  & $0.6451$   & \\
 $c_{\sigma K\bar{K}}$(fm)  &  -       &  & $1.0398$   & \\
 $G_{K\bar{K},\;K\bar{K}}$  &  -       &  & $0.0200$   & \\
 $d_{K\bar{K}}$(fm)         &  -       &  & $0.1000$   & \\
 $G_{\pi\pi,\;K\bar{K}}$    &  -       &  & $0.3500$   & \\
\hline\end{tabular}  \label{tab:expfit}
\end{center}
\end{table}

\subsection{Finite-volume Hamiltonian}
\label{sec:fvH1}
The finite-volume Hamiltonian method provides direct access to the
multi-particle energy eigenstates in a periodic volume
characterised by side length $L$.
The quantised three momenta of the $\pi$ meson
must be $k_n = \sqrt{n}\frac{2\pi}{L}$ for
integers $n = 0,1,2,\ldots$.
For a given choice of $N$ momenta $(k_0, k_1,\ldots, k_{N-1})$,
solving the Schrodinger equation $H|\Psi_E\rangle = E|\Psi_E\rangle $
in the finite box is equivalent to finding the solutions of the
following matrix equations
\begin{eqnarray}
\det \left([H_0]_{N+1} + [H_I]_{N+1} - E[I]_{N+1}\right) = 0
\label{eq:det}
\end{eqnarray}
where $det$ is taking the determinant of a matrix, $[I]_{N+1}$ is an
$(N+1)\times (N+1)$ unit matrix, and the non-interaction Hamiltonian
$H_0$, defined by Eq.(\ref{eq:h0}), is represented by the following
$(N+1)\times (N+1)$ matrix
\begin{eqnarray}
[H_0]_{N+1}&=&\left( \begin{array}{cccccc}
m_\sigma                & 0                             & 0                      & \cdots \\
0                       & 2\sqrt{k^2_0+m^2_{\pi}}       & 0                      & \cdots \\
0                       & 0                            & 2\sqrt{k^2_1+m^2_{\pi}} & \cdots \\
\vdots                  & \vdots                       & \vdots                 & \ddots
\end{array} \right),
\label{eq:h01}
\end{eqnarray}
With the forms of the interactions $g$ and $v$ in Eqs.(\ref{eq:int-g})-(\ref{eq:int-v}),
the $(N+1)\times (N+1)$ matrix representing
the interaction Hamiltonian $H_I$ can be written as
\begin{eqnarray}
[H_I]_{N+1}&=&\left( \begin{array}{cccccccc}
0                       & g^{fin}_{\pi\pi}(k_0)               & g^{fin}_{\pi\pi}(k_1)          & \cdots \\
g^{fin}_{\pi\pi}(k_0)   & v^{fin}_{\pi\pi,\pi\pi}(k_0, k_0)   & v^{fin}_{\pi\pi,\pi\pi}(k_0, k_1)  & \cdots \\
g^{fin}_{\pi\pi}(k_1)   & v^{fin}_{\pi\pi,\pi\pi}(k_1, k_0)   & v^{fin}_{\pi\pi,\pi\pi}(k_1, k_1)  & \cdots \\
\vdots                       & \vdots                                   & \vdots           & \ddots
\end{array} \right).
\label{eq:hi1}
\end{eqnarray}
The corresponding finite-volume matrix elements are given by
\begin{eqnarray}
g^{fin}_{\pi\pi}(k_n)&=&\sqrt{\frac{C_3(n)}{4\pi}}\left(\frac{2\pi}{L}\right)^{3/2}
g_{\sigma,\pi\pi}(k_n),\label{eq:gfin}\\
v^{fin}_{\pi\pi,\pi\pi}(k_{n_1},k_{n_2})&=&\sqrt{\frac{C_3(n_1)}{4\pi}}\sqrt{\frac{C_3(n_2)}{4\pi}}\left(\frac{2\pi}{L}\right)^3 v_{\pi\pi,\pi\pi}(k_{n_1},k_{n_2}),
\label{eq:vfin}
\end{eqnarray}
where $g_{\pi\pi}(k_n)$ and $v_{\pi\pi,\pi\pi}(k_{n_1},k_{n_2})$ are
defined in Eqs.(\ref{eq:gepp1})-(\ref{eq:vpp1}), and $C_3(n)$
represents the number of ways of summing the squares of three integers
to equal $n$.  As explained in Ref.\cite{Hall:2013qba}, the factor
$\sqrt{\frac{C_3(n)}{4\pi}}\left(\frac{2\pi}{L}\right)^{3/2}$ follows from
the quantization conditions in a finite box with size $L$.

The solution of Eq.(\ref{eq:det}) is a spectrum which depends on
the choice of the box size $L$ and $N$. Obviously, the acceptable solution must converge as
$N$ increases.
To get high accuracy results for examining L\"uscher's formula, we find that
 $N=600$ is sufficient for a range of $L$ in our calculations.
 The predicted spectra for each $L$ can be read from the solid curves
shown in Fig.~\ref{fg:com1}.
The dashed curves indicate the free-particle spectra (ie.~in the
absence of interactions).
%
  In a practical simulation at the physical pion mass, we note the
  energy threshold associated with the 4$\pi$ inelasticity is at $\sim
  560\mev$. The complete interpretation of energy levels near or above
  this threshold will necessarily involve new techniques which have
  yet to be developed. In this exploratory study, rather than going to
  a set of unphysical parameters or studying a toy model, we opt to
  study a realistic representation of the QCD interactions and neglect
  the role of multi-particle thresholds. For recent work on the
  extension to three-particle thresholds, the reader is referred to
  Refs.~\cite{Roca:2012rx,Polejaeva:2012ut,Kreuzer:2012sr,Briceno:2012rv,Hansen:2013dla,Guo:2013qla}.

\subsection{Phase shift extraction}
\label{sIIIc}
%

\begin{figure}[htbp] \vspace{-0.cm}
\begin{center}
\includegraphics[width=0.49\columnwidth]{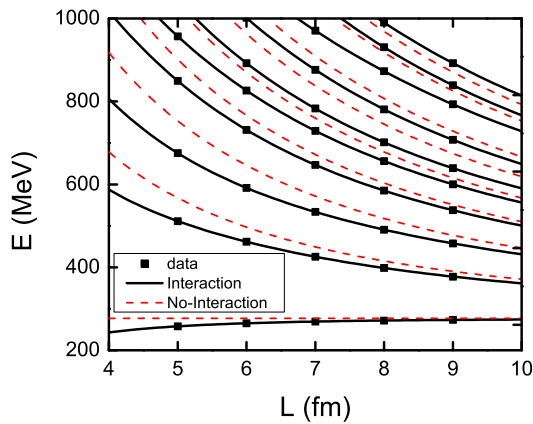}
\caption{(Color online) The spectrum of $\pi\pi$ states in the
  $1b-1c$ model.  The black curves are calculated by using the
  finite-volume Hamiltonian approach. The boxes denote discrete points
  on these curves which are used in the phase extraction shown in
  Fig.~\ref{fg:phase1c}.}
\label{fg:com1}
\end{center}
\end{figure}

\begin{figure}[htbp] \vspace{-0.cm}
\begin{center}
\includegraphics[width=0.49\columnwidth]{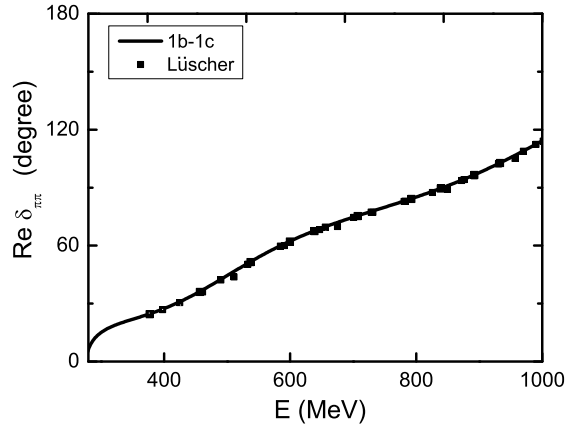}
\caption{The black curve is generated from directly solving
  scattering equations Eqs.(\ref{eq:lseq-1})-(\ref{eq:lseq-2}), and
  the solid squares are calculated from using the L\"uscher's method
  by using the spectrum appearing in Fig.~\ref{fg:com1}.}
\label{fg:phase1c}
\end{center}
\end{figure}

As reported in Ref.~\cite{Hall:2013qba}, the Hamiltonian and L\"uscher methods
predict almost identical finite volume spectra. The relationship between the Hamiltonian and L\"uscher quantisation conditions is explored analytically in Appendix~\ref{app:relationship}. Here we numerically demonstrate
this by using the L\"uscher formalism to extract the phase shift
from the finite volume spectra. The appropriate formulae are
summarised in Appendix~\ref{app:lusch}. By sampling the spectrum at a
discrete set of hypothetical volumes, shown in Fig.~\ref{fg:com1}, we
invert to obtain the phase shifts shown in Fig.~\ref{fg:phase1c}. Here we
see an excellent reproduction of the model phase shifts. A couple of
points show a small deviation from the exact curve. These correspond
to the smallest volume, $L=5\fm$, where the exponentially supressed
corrections are beginning to be relevant.

In comparison with realistic lattice calculations, we note that the
smooth reproduction of the phase shift would require significant
resources in terms of the number of volumes sampled. Such a dense extraction of
the phase shift is more easily made possible by studying the spectra
in moving frames, such as
Ref.~\cite{Dudek:2012xn, Rummukainen:1995vs, Kim:2005gf, Fu:2011xz, Leskovec:2012gb,Gockeler:2012yj}.
The extension of
the Hamiltonian formalism to such boosted systems will be
investigated in future work.

With the equivalence with the L\"uscher technique demonstrated, we
now turn to the extension to multi-channel scattering.

\section{one bare state and two-channels}

\subsection{Model parameters}
\begin{figure}[htbp] \vspace{-0.cm}
\begin{center}
\includegraphics[width=0.49\columnwidth]{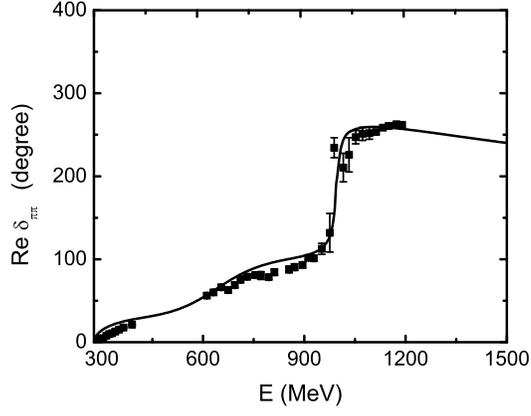}
\caption{The phase shift $\delta_{\pi\pi}$
 for $\pi\pi$ scattering from the  $1b-2c$ model
are compared with the data.}
\label{fg:expfit2pipi}
\end{center}
\end{figure}

\begin{figure}[htbp] \vspace{-0.cm}
\begin{center}
\includegraphics[width=0.49\columnwidth]{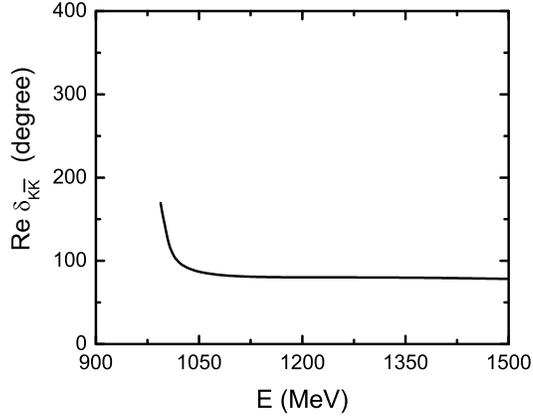}
\caption{The phase shift $\delta_{K\bar{K}}$
 of $K\bar{K}$ scattering calculated in the $1b-2c$ model.}
\label{fg:expfit2KK}
\end{center}
\end{figure}

\begin{figure}[htbp] \vspace{-0.cm}
\begin{center}
\includegraphics[width=0.49\columnwidth]{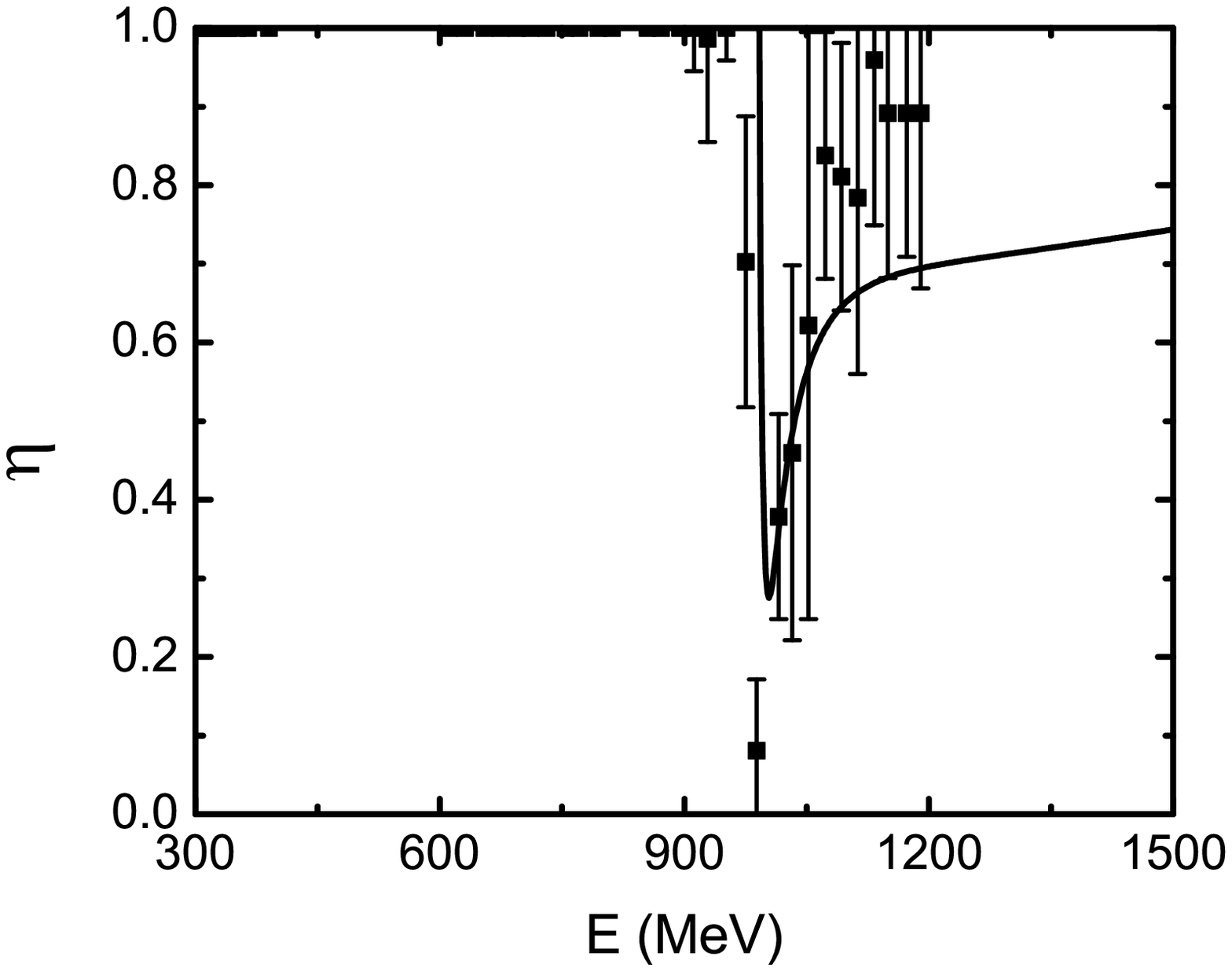}
\caption{The inelasticity $\eta$ in the $1b-2c$ model
compared with the data.}
\label{fg:expfit2eta}
\end{center}
\end{figure}

To describe $\pi\pi$ scattering above the $K\bar{K}$ threshold, we
construct a model with one bare state and two-channels. The formula
for such a model can be obtained from Section II by setting $n=1$ for
a bare particle $\sigma$ and $\alpha, \beta,\gamma =\pi\pi,
K\bar{K}$. Similar to the $1b-1c$ model of section III, the matrix
elements of the interactions defined in Eqs.(\ref{eq:int-g}) and
(\ref{eq:int-v}) are parameterized as
\begin{eqnarray}
\langle \sigma|g| \alpha(\vec{k})\rangle &=&g_{\sigma,\alpha}(k)\nonumber \\
&=&\frac{g_{\sigma,\alpha}}{\sqrt{\pi}}\frac{1}{(1+(c_{\alpha}\times k)^2)},\label{eq:g2}\\
\langle \alpha(\vec{k})|v|\beta(\vec{k}')\rangle  &=& v_{\alpha, \beta}(k,k')\nonumber \\
&=&\frac{G_{\alpha,\;\beta}}{m^2_\pi}\times\frac{1}{(1+(d_{\alpha}\times k)^2)^2}
\times\frac{1}{(1+(d_{\beta}\times k')^2)^2},\label{eq:v2}
\end{eqnarray}
%
with $k$ and $k'$ are the three momenta of $\pi$ or $K$ in the center
mass system.
There are ten parameters: $m_\sigma$, $g_{\pi\pi}$,
$c_{\pi\pi}$, $g_{K\bar{K}}$,
$c_{K\bar{K}}$,
$G_{\pi\pi,\;\pi\pi}$, $G_{\pi\pi,\;K\bar{K}}$, $G_{K\bar{K},\;K\bar{K}}$
$d_{\pi\pi}$ and $d_{K \bar{K}}$.
By fitting the data of $\pi\pi$ phase shift $\delta$ and inelasticity $\eta$,
the model parameters can be determined and are listed in the second column of
 Table~\ref{tab:expfit}.
 The calculated phase shifts are compared with the data in
 Figs.~\ref{fg:expfit2pipi}--\ref{fg:expfit2eta}.  As in the single
 channel case, the agreement is sufficiently good for our exploration
 of the finite volume Hamiltonian method.

\subsection{Finite-volume Hamiltonian method}
\begin{figure}[htbp] \vspace{-0.cm}
\begin{center}
\includegraphics[width=0.49\columnwidth]{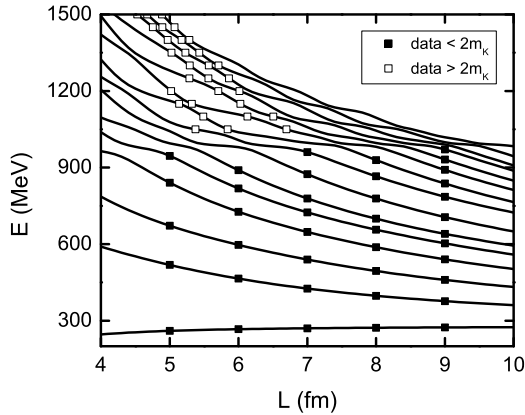}
\caption{
 The black curves show the energy
spectra generated by using
the finite-box Hamiltonian approach within the the $1b-2c$ model.
The solid and open squares are selected solutions below and above the inelastic threshold, respectively.
These solutions have been inverted through the extended L\"uscher
formalism to determine the phase shifts and inelasticities, see Fig.~\ref{fg:com2}.}
\label{fg:spectra2c}
\end{center}
\end{figure}

To calculate the spectrum for the $1b-2c$ model constructed in the
previous subsection, we follow the procedures given in Section
\ref{sec:fvH1} to extend the matrix representation of the Hamiltonian
to include the elements associated with the additional $K\bar{K}$
channel for each mesh points of the chosen $N$ momenta $k_n =
\sqrt{n}\frac{2\pi}{L}$ for $ n = 0, 1,2\cdot\cdot\cdot (N-1)$.  This
leads to the following $(2N+1)\times (2N+1)$ matrix equations
\begin{eqnarray}
\det ([H_0]_{2N+1} + [H_I]_{2N+1} - E[I]_{2N+1}) = 0
\label{eq:det-1}
\end{eqnarray}
where $[I]_{2N+1}$ is an $(2N+1)\times(2N+1)$ unit matrix, and
\begin{eqnarray}
[H_0]_{2N+1}&=&\left( \begin{array}{cccccccc}
m_0                     & 0                            & 0
                        & 0                            & 0  & \cdots \\
0                       & 2\sqrt{k^2_0+m^2_{\pi}}      & 0
                        & 0                            & 0  & \cdots \\
0                       & 0                            & 2\sqrt{k^2_0+m^2_{K}}
                        & 0                            & 0  & \cdots \\
0                       & 0                            & 0
                        & 2\sqrt{k^2_1+m^2_{\pi}}      & 0  & \cdots \\
0                       & 0                            & 0
                        & 0                            & 2\sqrt{k^2_1+m^2_{K}}  & \cdots \\
\vdots                  & \vdots                       & \vdots
                        & \vdots                       & \vdots                & \ddots
\end{array} \right) \nonumber
\end{eqnarray}
The $(2N+1)\times(2N+1)$ matrix for the interaction Hamiltonian is
\begin{eqnarray}
[H_I]_{2N+1}&=&\left( \begin{array}{cccccccc}
0                       & g^{fin}_{\pi\pi}(k_0)               & g^{fin}_{K\bar{K}}(k_0)
                        & g^{fin}_{\pi\pi}(k_1)               & g^{fin}_{K\bar{K}}(k_1)              & \cdots \\
g^{fin}_{\pi\pi}(k_0)   & v^{fin}_{\pi\pi,\pi\pi}(k_0, k_0)   & v^{fin}_{\pi\pi,K\bar{K}}(k_0, k_0)
                        & v^{fin}_{\pi\pi,\pi\pi}(k_0, k_1)   & v^{fin}_{\pi\pi,K\bar{K}}(k_0, k_1)   & \cdots \\
g^{fin}_{K\bar{K}}(k_0) & v^{fin}_{K\bar{K},\pi\pi}(k_0, k_0) & v^{fin}_{K\bar{K},K\bar{K}}(k_0, k_0)
                        & v^{fin}_{K\bar{K},\pi\pi}(k_0, k_1) & v^{fin}_{K\bar{K},K\bar{K}}(k_0, k_1)   & \cdots \\
g^{fin}_{\pi\pi}(k_1)   & v^{fin}_{\pi\pi,\pi\pi}(k_1, k_0)   & v^{fin}_{\pi\pi,K\bar{K}}(k_1, k_0)
                        & v^{fin}_{\pi\pi,\pi\pi}(k_1, k_1)   & v^{fin}_{\pi\pi,K\bar{K}}(k_1, k_1)   & \cdots \\
g^{fin}_{K\bar{K}}(k_1) & v^{fin}_{K\bar{K},\pi\pi}(k_1, k_0) & v^{fin}_{K\bar{K},K\bar{K}}(k_1, k_0)
                        & v^{fin}_{K\bar{K},\pi\pi}(k_1, k_1) & v^{fin}_{K\bar{K},K\bar{K}}(k_1, k_1)   & \cdots \\
\vdots                  & \vdots                       & \vdots
                        & \vdots                       & \vdots                  & \ddots
\end{array} \right), \nonumber
\end{eqnarray}
with
\begin{eqnarray}
g^{fin}_{\alpha}(k_n)&=&\sqrt{\frac{C_3(n)}{4\pi}}\left(\frac{2\pi}{L}\right)^{3/2}
g_{\sigma,\alpha}(k_n),\\
v^{fin}_{\alpha,\beta}(k_{n_i},k_{n_j})&=&\sqrt{\frac{C_3(n_i)}{4\pi}}
\sqrt{\frac{C_3(n_j)}{4\pi}}\left(\frac{2\pi}{L}\right)^3 v_{\alpha,\beta}(k_{n_i},k_{n_j}),
\label{eq:vfin-1}
\end{eqnarray}
where $g_{\sigma,\alpha}(k_n)$ and $v_{\alpha,\beta}(k_{n_i},k_{n_j})$
are defined in Eqs.~(\ref{eq:g2}) and (\ref{eq:v2}). In this way,
we can generate the spectrum from the Hamiltonian in a finite box
with a given size $L$ by solving Eq.~(\ref{eq:det-1}). The computed
spectrum is shown as a function of the volume in
Fig.~\ref{fg:spectra2c}.

   As discussed in the previous section, we are neglecting the
  physics associated with the multiparticle thresholds (eg. $4\pi$ at
  $E\sim 560\mev$). We thereby focus our attention on the issues
  related to the coupled-channel system, while maintaining a realistic
  representation of observed scattering in QCD.

\subsection{Multi-channel spectra}
\label{sIVc}
Our first task here is to establish the equivalence of the Hamiltonian
spectrum with that of the multi-channel generalisation of
L\"uscher. The relevant formulae for the coupled-channel system are
summarised in Appendix~\ref{app:multi}. For the present case, the
eigenvalue spectrum (above the inelastic threshold) is defined by the
solutions to the following equation
\begin{align}
&\cos\left[\phi(q_{\pi\pi})+\phi(q_\KKbar)-\delta_{\pi\pi}(E)-\delta_{K\bar{K}}(E)\right]\non\\
&\quad-\eta(E)\cos\left[\phi(q_{\pi\pi})-\phi(q_\KKbar)-\delta_{\pi\pi}(E)+\delta_{K\bar{K}}(E)\right]=0.
\label{eq:ev2c}
\end{align}
The phase $\phi$ characterises the lattice geometry as defined by
Eq.~(\ref{eq:phi}). Knowledge of the energy-dependence of the phase
shifts and inelasticity allows one to determine the spectrum for a
given value of $L$.
The eigenvalue equation is solved for $E$, where
the dimensionless momenta, $q_\alpha=k_\alpha L/(2\pi)$, corresponding
to the on-shell momentum $k_\alpha$ in channel $\alpha$ (see
Eq.~(\ref{eq:onshell})).

\begin{figure}[htbp] \vspace{-0.cm}
\begin{center}
\includegraphics[width=0.49\columnwidth]{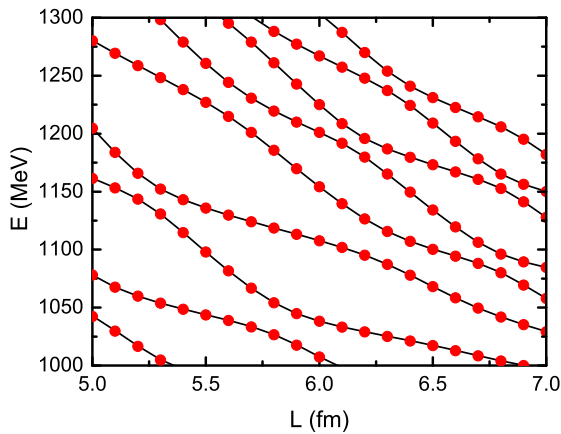}
\caption{(Color online) The solid dots represent the finite volume spectrum as determined by the extended L\"uscher formalism; computed directly from the model phase shifts and inelasticities. These are in excellent agreement with the spectra computed with the Hamiltonian approach, as shown by the continuous curves.
}
\label{fg:multicomp}
\end{center}
\end{figure}

Using the model phase shifts and inelasticities, the L\"uscher-style
formalism allows one to uniquely determine the finite volume
spectrum. For this model, the solutions of Eq.~(\ref{eq:ev2c}) (in the
inelastic region) are shown in Fig.~\ref{fg:multicomp}. The predicted
spectra within the two approaches are in excellent agreement
---
  hence confirming that the spectra are determined by the same
  asymptotic eigenvalue constraint.
%



Of relevance to lattice QCD simulations is the desire to obtain
$\delta_{\pi\pi}$, $\delta_{K\bar{K}}$ and $\eta$ from the spectra
determined in numerical simulations.  Using Eq.~(\ref{eq:luch2a}), the
isolation of all three scattering parameters at any given $E$ would
require eigenstates at this energy for three different box
sizes.\footnote{
    Of course in any finite statistics simulation,
    this degeneracy will only be realised up to some finite numerical
    precision.
    } Such solutions are indicated by the white squares in
Fig.~\ref{fg:spectra2c}.
%
%
  Across an ensemble of volumes, the extraction of the resonance
  parameters from the asymptotic constraints of the L\"uscher
  quantisation alone, can only lead to a ``pointwise'' determination
  of the scattering parameters. Such a ``pointwise'' inversion for the
  coupled-channel systems was discussed by Guo et
  al.~\cite{Guo:2012hv}. Here it was demonstrated that by using
  multiple different total momentum quantisations of the system, there
  is an increased opportunity to identify near-degenerate eigenstates
  such that at least three independent qualisations can be used to
  model-independently extract the scattering parameters. Nevertheless,
  it is generally true for any finite set of discrete spectra, the
  pointwise extraction will only have a limited applicability.

\begin{figure}[htbp] \vspace{-0.cm}
\begin{center}
\includegraphics[width=0.49\columnwidth]{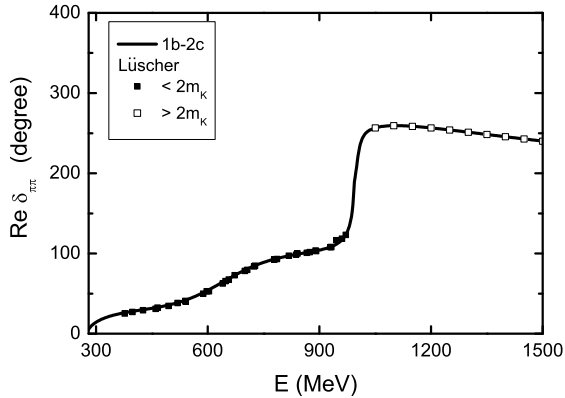}
\caption{
The black curve denotes the model $\pi\pi$ phase shift.
The solid and open squares denote the inversion of the solutions shown in Fig.~\ref{fg:spectra2c} below and above the inelastic thresholds.
Below the inelastic threshold, each solution uniquely determines the phase shift.
Above the inelastic threshold, the unique solution requires the impractical determination of three
identical energy levels at different $L$. In this region,
$\delta_\KKbar$ and $\eta$ (Figs.~\ref{fg:expfit2KK} and
\ref{fg:expfit2eta}) are equally-well described by this inversion.}
\label{fg:com2}
\end{center}
\end{figure}

For an example of the inversion in the present case,
 at
$E=1200\mev$, with box sizes $L=5.022,5.708,6.014\fm$, the model
spectrum can be inverted through Eq.~(\ref{eq:luch2a}) to determine
\begin{equation}
\delta_\pipi=256.5^\circ,\quad \delta_\KKbar=79.84^\circ,\quad \eta=0.6980.
\end{equation}
We note the relative phase between $\delta_\pipi$ and $\delta_\KKbar$
is only determined up to integer multiples of $\pi$ --- an ambiguity
that has been elaborated on in Ref.~\cite{Berkowitz:2012xq}.  Up to
the determination of this phase, we note excellent agreement with the
underlying model scattering,
\begin{equation}
\delta_\pipi=256.6^\circ,\quad \delta_\KKbar=80.18^\circ,\quad \eta=0.6965.
\end{equation}
The extraction of $\delta_\pipi$ in this way, for a range of energies,
is shown by the white squares in Fig.~\ref{fg:com2}.

  To make the most of a finite set of spectrum ``data'',
  Ref.~\cite{Guo:2012hv} have proposed using a $K$-matrix formulation
  to parameterise the $S$-matrix and thereby the predicted spectrum.
  In the following Section we explore the use of the Hamiltonian
  formulation as an alternative parameterisation to fit a finite set
  of lattice spectra. Both the Hamiltonian and $K$-matrix approaches
  have been used extensively to extract from scattering observables
  the resonance parameters associated with the excited hadrons; as
  reviewed in Ref.~\cite{Burkert:2004sk} for the excited nucleons. It
  has been well recognised that the comparisons of the results from
  these two different approaches are fruitful in making progress to
  establish the hadron spectra; as can be seen in the coupled-channel
  analysis results presented in
  Refs.~\cite{Kamano:2013iva,Anisovich:2011fc,Ronchen:2012eg}.

We note that the main point of our approach is to relate the spectrum in a
finite volume to the asymptotic properties of scattering wavefunctions directly
through a procedure of diagonalizing a Hamiltonian; rather than indirectly through
the scattering parameters. Our numerical results presented above show that this
procedure is equivalent to the L\"uscher formulation for the coupled-channel case.
Thus our approach is readily applicable to the case with more than two particles,
for which the corresponding L\"uscher formulation has not yet been developed. This
marks the main difference between our work with that of Ref.~\cite{Guo:2012hv}, and similarly
related work.


\section{Applications to LQCD}
We investigate the procedure for
using the Hamiltonian approach to
predict the scattering observables from the spectrum generated from LQCD.
We will compare our approach with the approach based on L\"uscher's formula.
For this illustrative purpose, it is sufficient to  use the  $1b-1c$ and $1b-2c$ models
described in
sections III and IV to generate the spectra
which will be referred to
as the ``LQCD data''. The phase shifts at each energy of
these spectra are of course known, as shown as the solid
curves in Figs.~\ref{fg:expfit1} and \ref{fg:expfit2pipi}--\ref{fg:expfit2eta}.

Our procedure is to use a Hamiltonian to fit a given choice of the
spectrum data by solving the eigenvalue problem defined by
Eqs.(\ref{eq:det})-(\ref{eq:vfin}) for the one-channel case and
Eqs.(\ref{eq:det-1})-(\ref{eq:vfin-1}) for the two-channel case.  We
then use the determined Hamiltonian to calculate the phase shifts by
using the scattering equations Eqs.(\ref{eq:lseq-1})-(\ref{eq:lseq-2})
in infinite space.

To proceed, we need to choose the forms of the interactions in
Eqs.~(\ref{eq:h})--(\ref{eq:int-v}) of the phenomenological
Hamiltonian.  For simplicity, we consider the Hamiltonian which has
either one bare state and one-channel or one bare state and
two-channels. These Hamiltonians are similar to the $1b-1c$ and
$1b-2c$ models constructed in sections III and IV, but they can have a
different parametrization of the vertex interaction
$g_{\sigma,\alpha}$ and $v_{\alpha,\beta}$. We consider three forms:
\begin{itemize}
\item A:
\begin{eqnarray}
g(k)_{\sigma,\alpha}&=&\frac{g_{\alpha}}{\sqrt{\pi}}\frac{1}
{(1+(c_{\alpha}\times k)^2)},\label{eq:geqa}\\
v_{\alpha,\beta}(k,k')&=&\frac{G_{\alpha,\;\beta}}
{m^2_\pi}\times\frac{1}{(1+(d_{\alpha}\times k)^2)^2}
\times\frac{1}{(1+(d_{\beta}\times k')^2)^2},\label{eq:veqa}
\end{eqnarray}

\item B:
\begin{eqnarray}
g(k)_{\sigma,\alpha}&=&\frac{g_{\alpha}}{\sqrt{\pi}}\frac{1}
{(1+(c_{\alpha}\times k)^2)^2},\label{eq:geqb}\\
v_{\alpha,\beta}(k,k')&=&\frac{G_{\alpha,\;\beta}}
{m^2_\pi}\times\frac{1}{(1+(d_{\alpha}\times k)^2)^4}
\times\frac{1}{(1+(d_{\beta}\times k')^2)^4},\label{eq:veqb}
\end{eqnarray}

\item C:
\begin{eqnarray}
g(k)_{\sigma,\alpha}&=&\frac{g_{\alpha}}{\sqrt{\pi}}e^{-(c_{\alpha}\times k)^2},\label{eq:geqc}\\
v_{\alpha,\beta}(k,k')&=&\frac{G_{\alpha,\beta}}{m^2_\pi}
e^{-(d_{\alpha}\times k)^2} e^{-(d_{\beta}\times k')^2},\label{eq:veqc}
\end{eqnarray}

Note that the parametrization $A$ is the same as those of models $1b-1c$ and $1b-2c$, as described above.

\end{itemize}

\subsection{Fit for one-channel}
\begin{figure}[htbp] \vspace{-0.cm}
\begin{center}
\includegraphics[width=0.49\columnwidth]{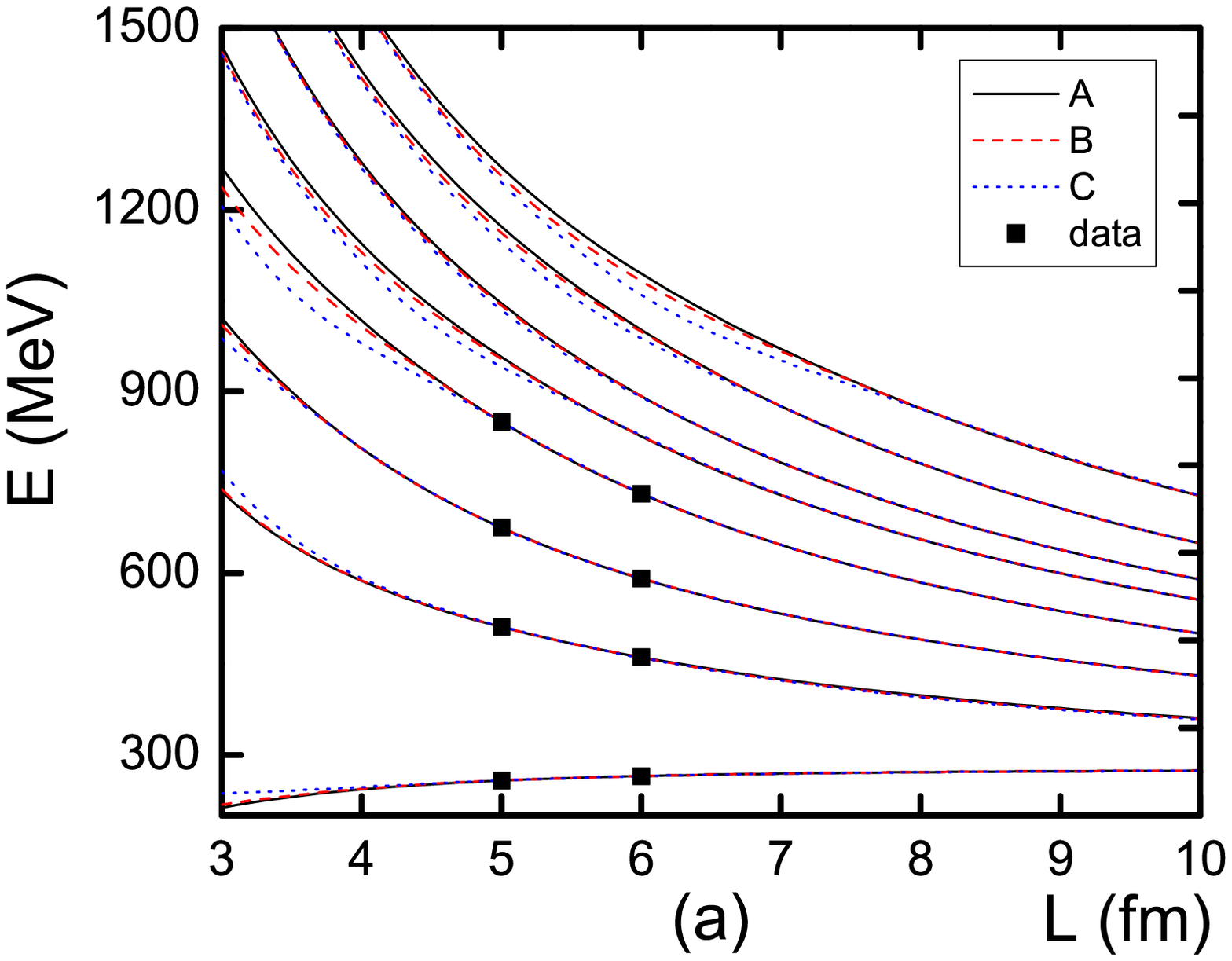}
\includegraphics[width=0.49\columnwidth]{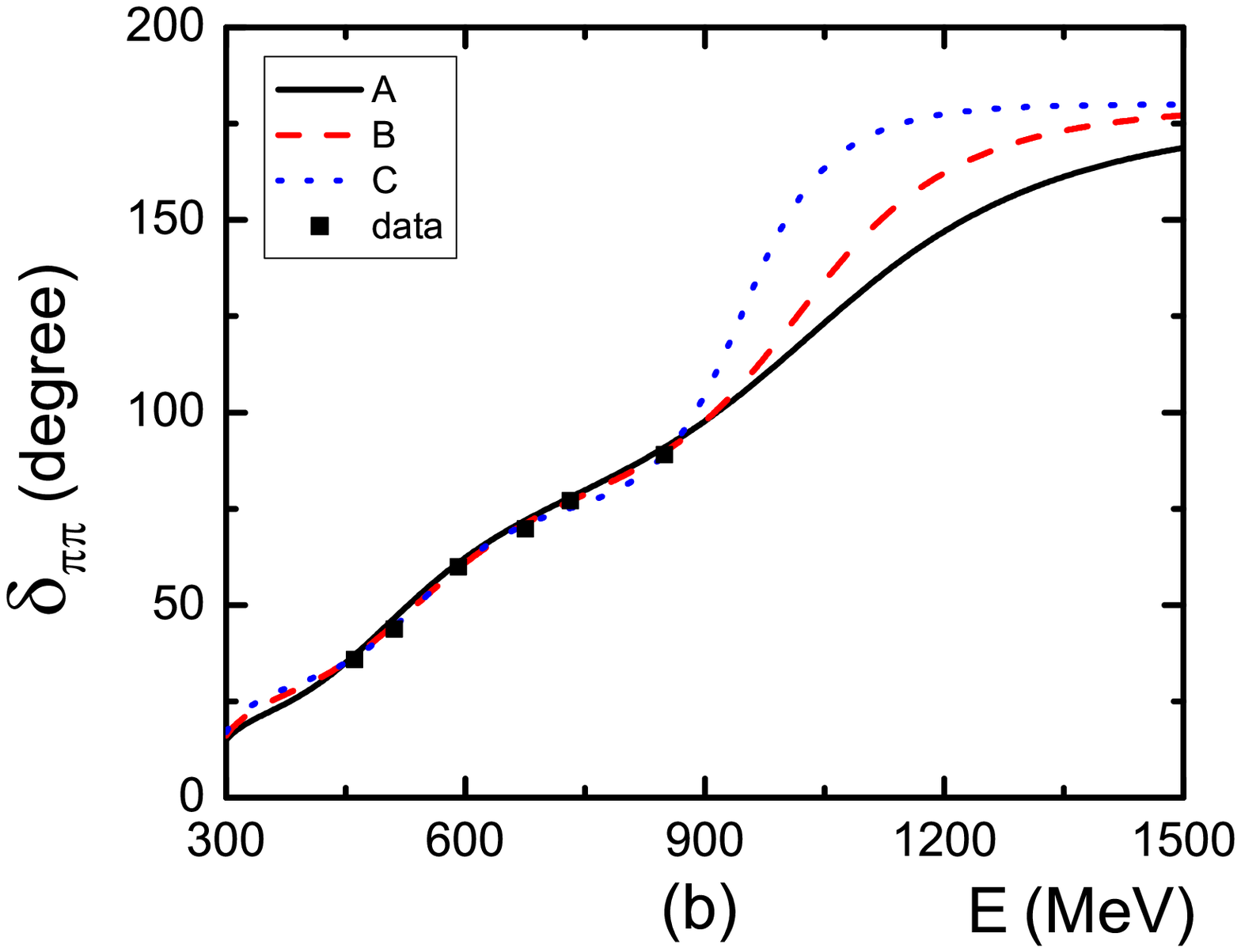}
\caption{(Color online) (a): the spectrum data generated from $1b-1c$ model. (b): the
phase shifts calculated from the one-channels model with
parametrization $A$ ($1b-1c$ model) and $B$ and $C$ specified
in Eqs.(\ref{eq:geqa})-(\ref{eq:veqc}) are compared with the data (from
$1b-2c$ model).} \label{fg:fit18}
\end{center}
\end{figure}

We first consider the one-channel case. The spectrum data are
generated from model $1b-1c$ constructed in section III.  In the left
side of Fig.~\ref{fg:fit18}, we show 8 data points generated by
solving the eigenvalue equation, Eq.~(\ref{eq:det}), for $L = 5, 6$ fm.
%
For the discussion of this manuscript, the choice of $L$ values
is largely irrelevant. The smaller of these volumes has $m_\pi L\sim
  3.5$, which is just below the reputed value of 4. As such, it is
  plausible that there are non-negligible corrections associated with
  the exponentially suppressed finite-volume effects
  \cite{Bedaque:2006yi,Chen:2012rp,Albaladejo:2013bra}. While we
  neglect these effects in the present study, they will certainly be
  of relevance in future precision studies.

To see whether the fit depends sensitively on the form of the
Hamiltonian, we assign a very small ($1$ MeV) error for each energy
level in the spectrum.
We find that these 8 spectrum data points can be fitted by
using the parametrization $B$, or $C$, as shown in the left side of
Fig.\ref{fg:fit18}.
%
The $\pi\pi$ phase
shifts calculated from two new Hamiltonians using the scattering
equations Eqs.~(\ref{eq:lseq-1})--(\ref{eq:lseq-2}) in infinite space
are compared with the data (solid squares) in the right side of
Fig.~\ref{fg:fit18}.  They agree very well in the energy region $E$
$\lesssim$ 0.9 GeV, where the spectrum data are fitted.  At higher
energies, the calculated phase shifts from $B$ and $C$ deviate from
each other and also from the $1b-1c$ model.  Note that both the black
solid curves ($A$) and data (solid squares) are from the $1b-1c$ model
and thus they agree with each other completely.

\begin{figure}[htbp] \vspace{-0.cm}
\begin{center}
\includegraphics[width=0.49\columnwidth]{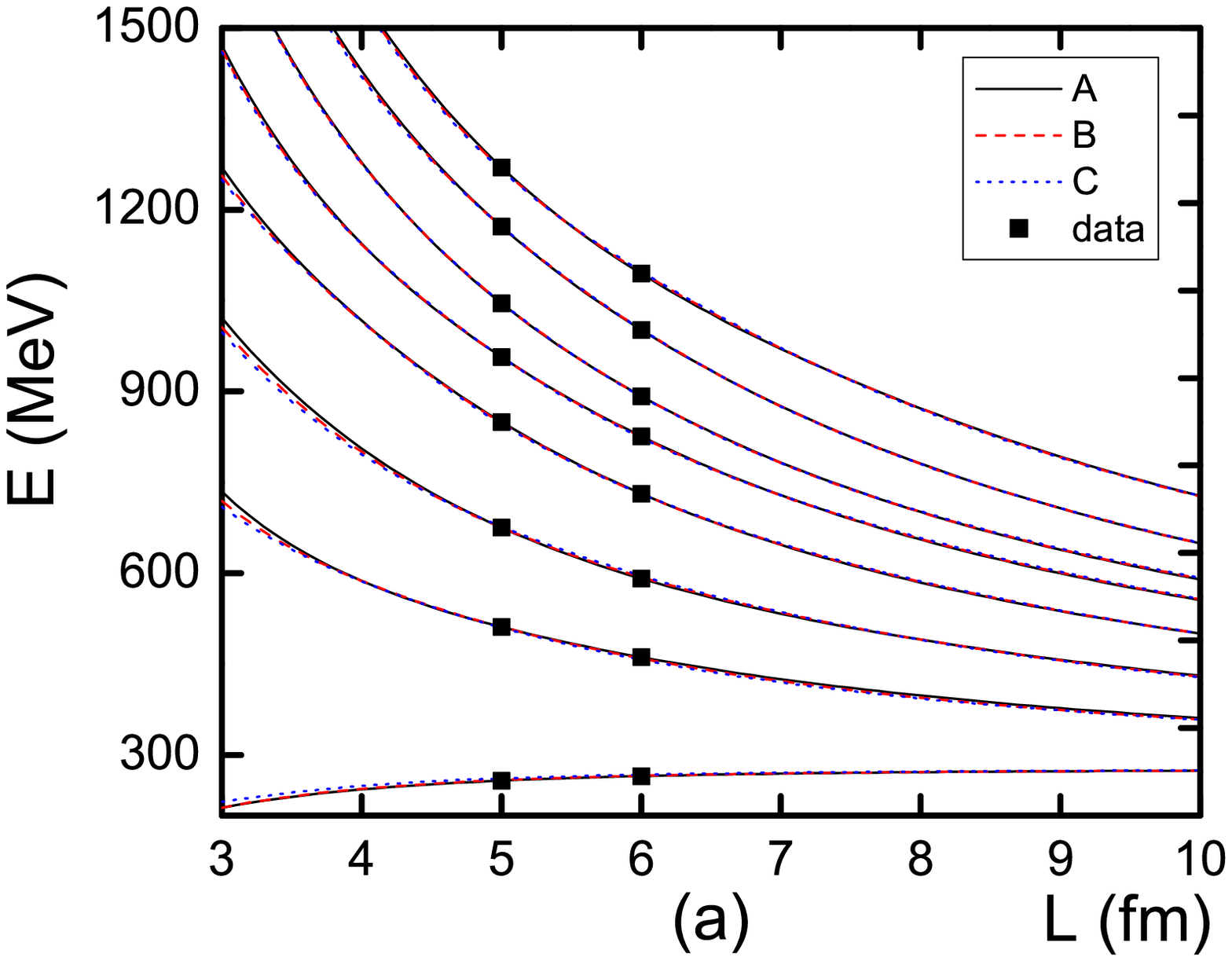}
\includegraphics[width=0.49\columnwidth]{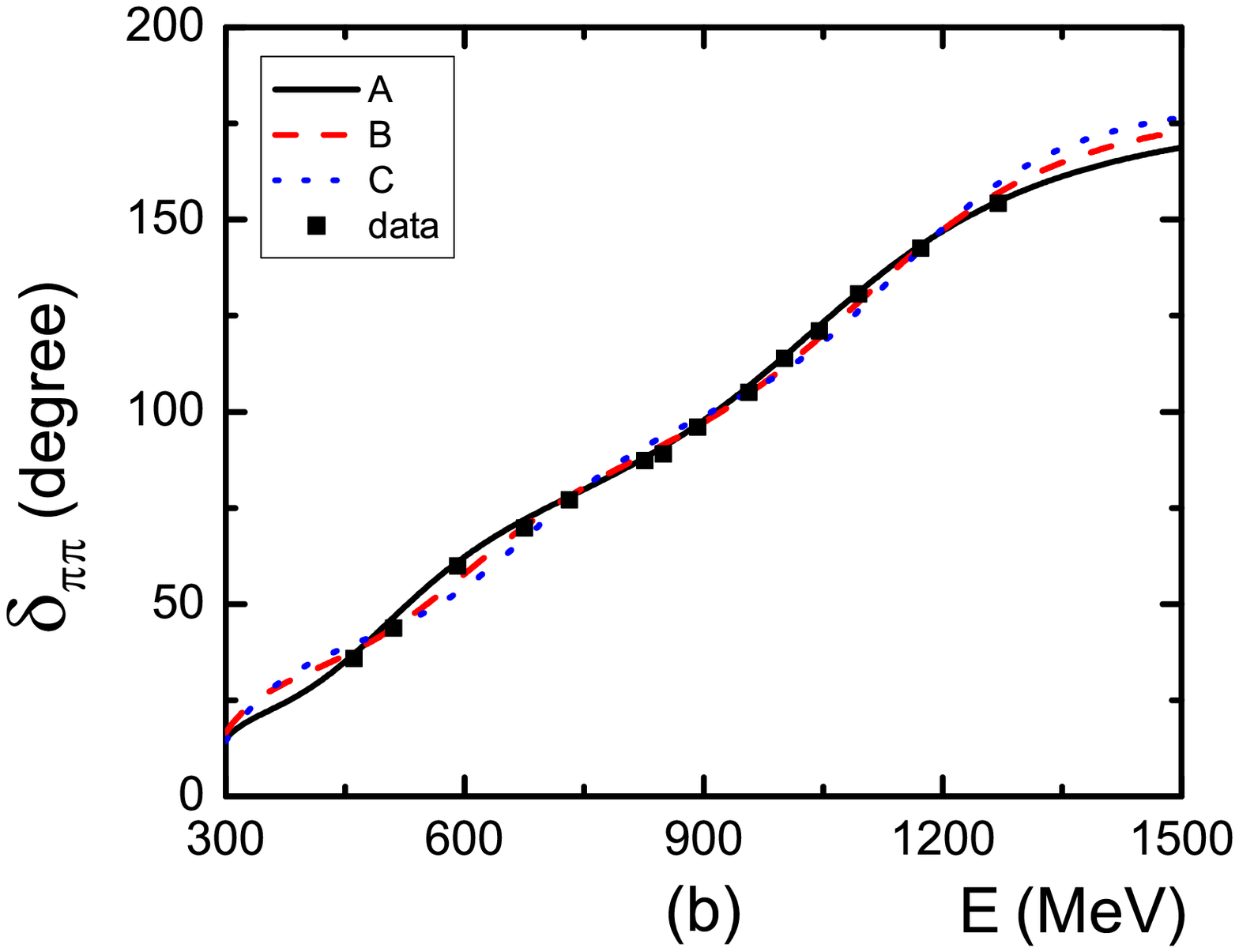}
\caption{(Color online) (a): the spectrum data generated from $1b-1c$ model. (b): the
phase shifts calculated from the one-channels model with
the parametrization $A$ ($1b-1c$ model), $B$ and $C$
specified in Eqs.(\ref{eq:geqa})-(\ref{eq:veqc}) are
compared with the data (from $1b-1c$ model).} \label{fg:fit116}
\end{center}
\end{figure}

The results presented above suggest that the finite-box Hamiltonian
approach is valid in the energy region where the spectrum data are
fitted, since the predicted scattering phase shifts are independent of
the form of the Hamiltonian and agree with the phase shifts
corresponding the fitted spectrum data.
To further examine this, we generate 16 data points up to 1.2 GeV and
repeat the fitting process. The generated data are the black squares
in the left side of Fig.\ref{fg:fit116}.
The predicted phase shifts agree with the data in the $E <$ 1.2 GeV
region where the spectrum data are fitted. Above 1.2 GeV they deviate
from the the $1b-1c$ model, similar to what we observed in
Fig.\ref{fg:fit18}.

With the results shown in Figs.~\ref{fg:fit18}--\ref{fg:fit116} and the
Fig.~\ref{fg:com1} on L\"uscher's method in section III, we
conclude that the finite-volume Hamiltonian approach gives a
comparable reproduction of the phase shifts as compared to L\"uscher's
method.
However, for the one-channel case the finite-volume Hamiltonian method
has no distinct advantage over L\"uscher's method, since the
required LQCD efforts are not so different.

\subsection{Fit for two-channels}
\begin{figure}[htbp] \vspace{-0.cm}
\begin{center}
\includegraphics[width=0.49\columnwidth]{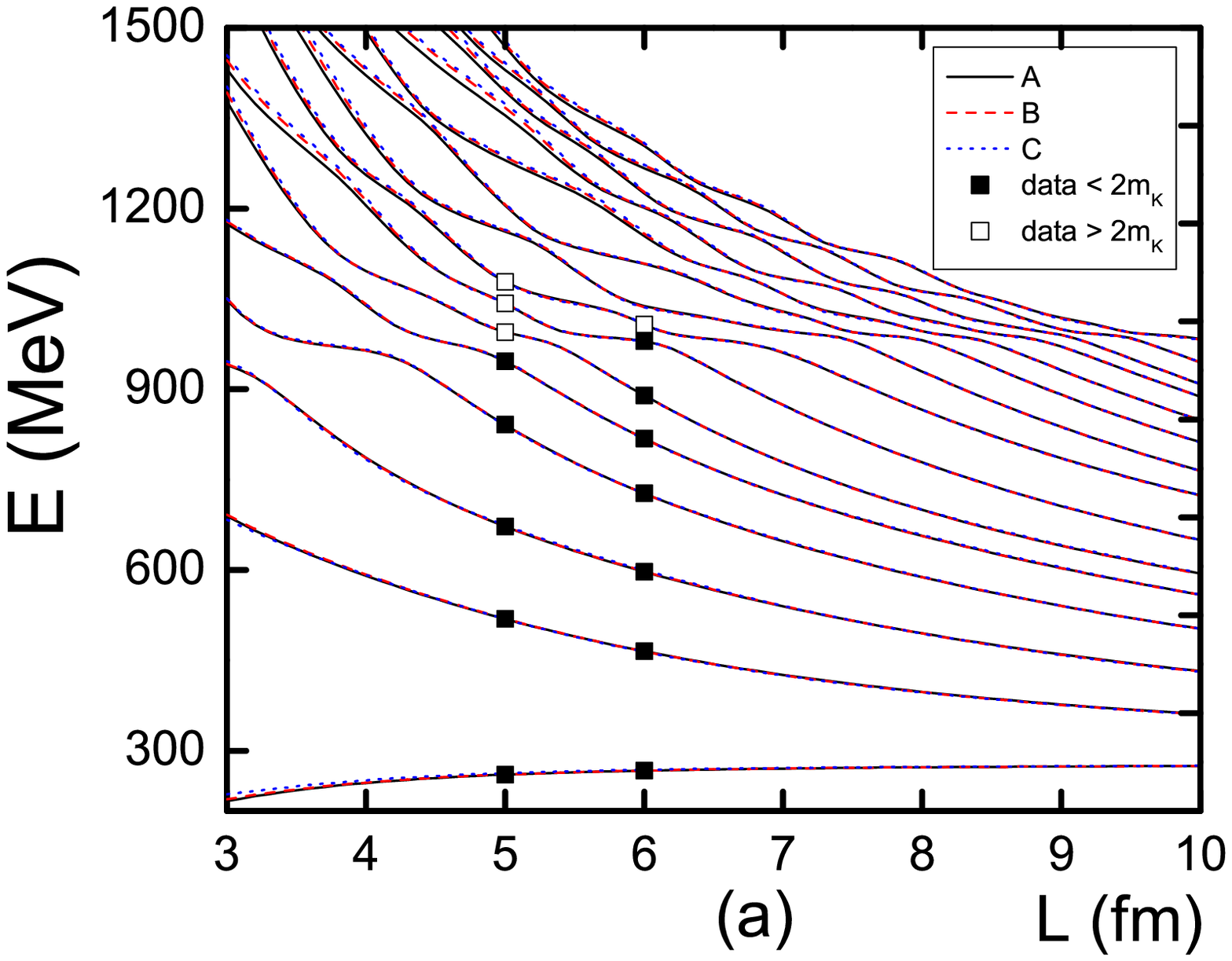}
\includegraphics[width=0.49\columnwidth]{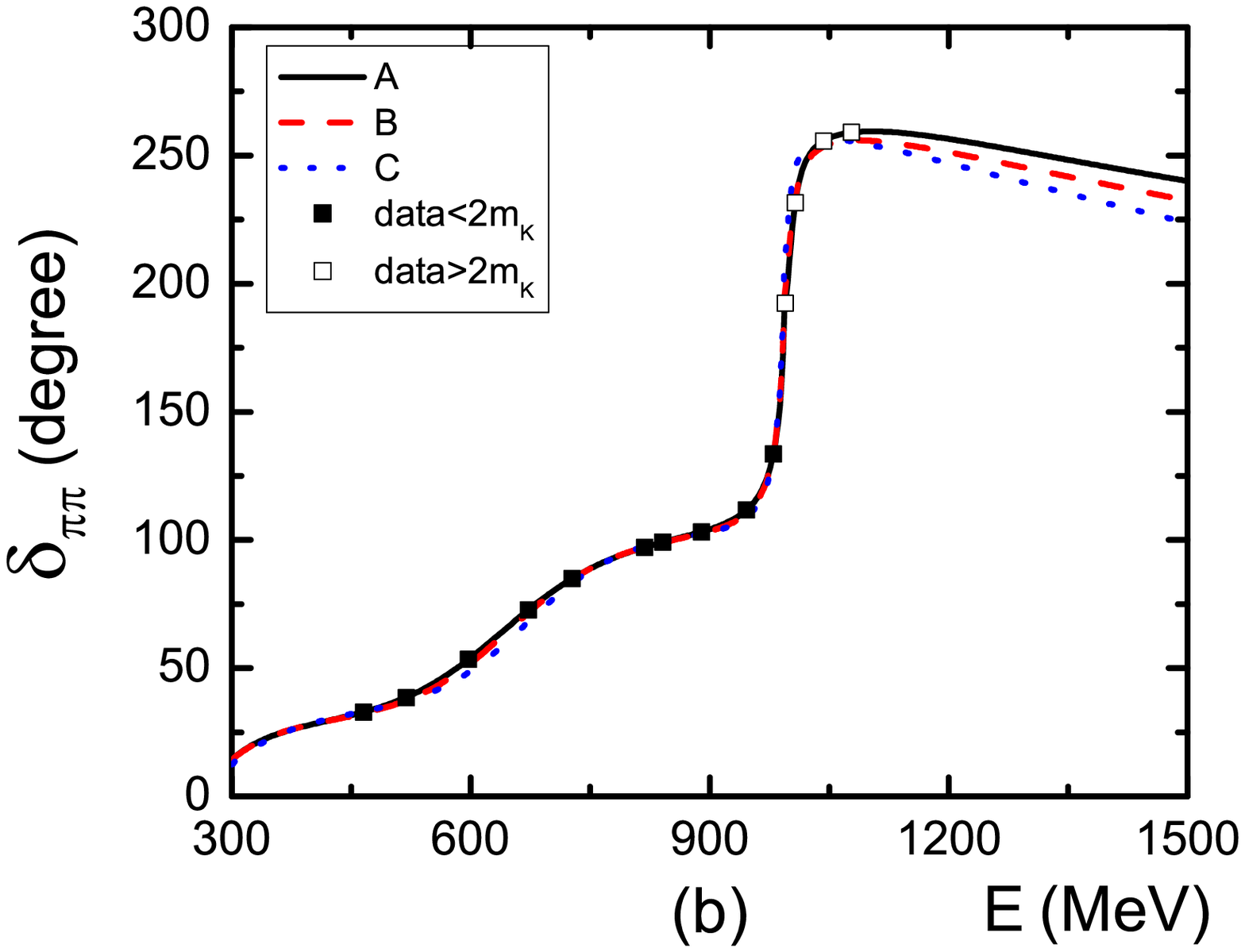}
\includegraphics[width=0.49\columnwidth]{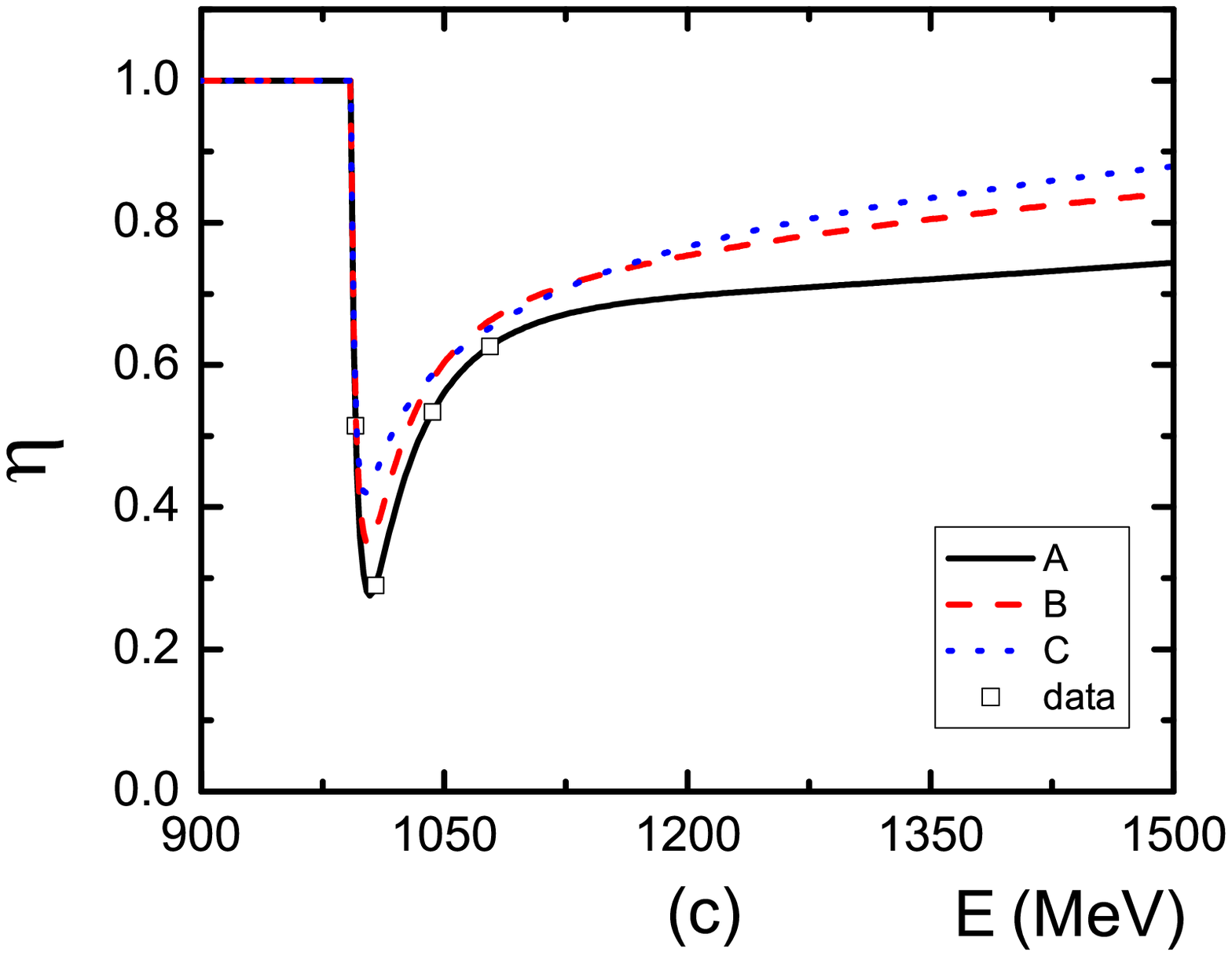}
\includegraphics[width=0.49\columnwidth]{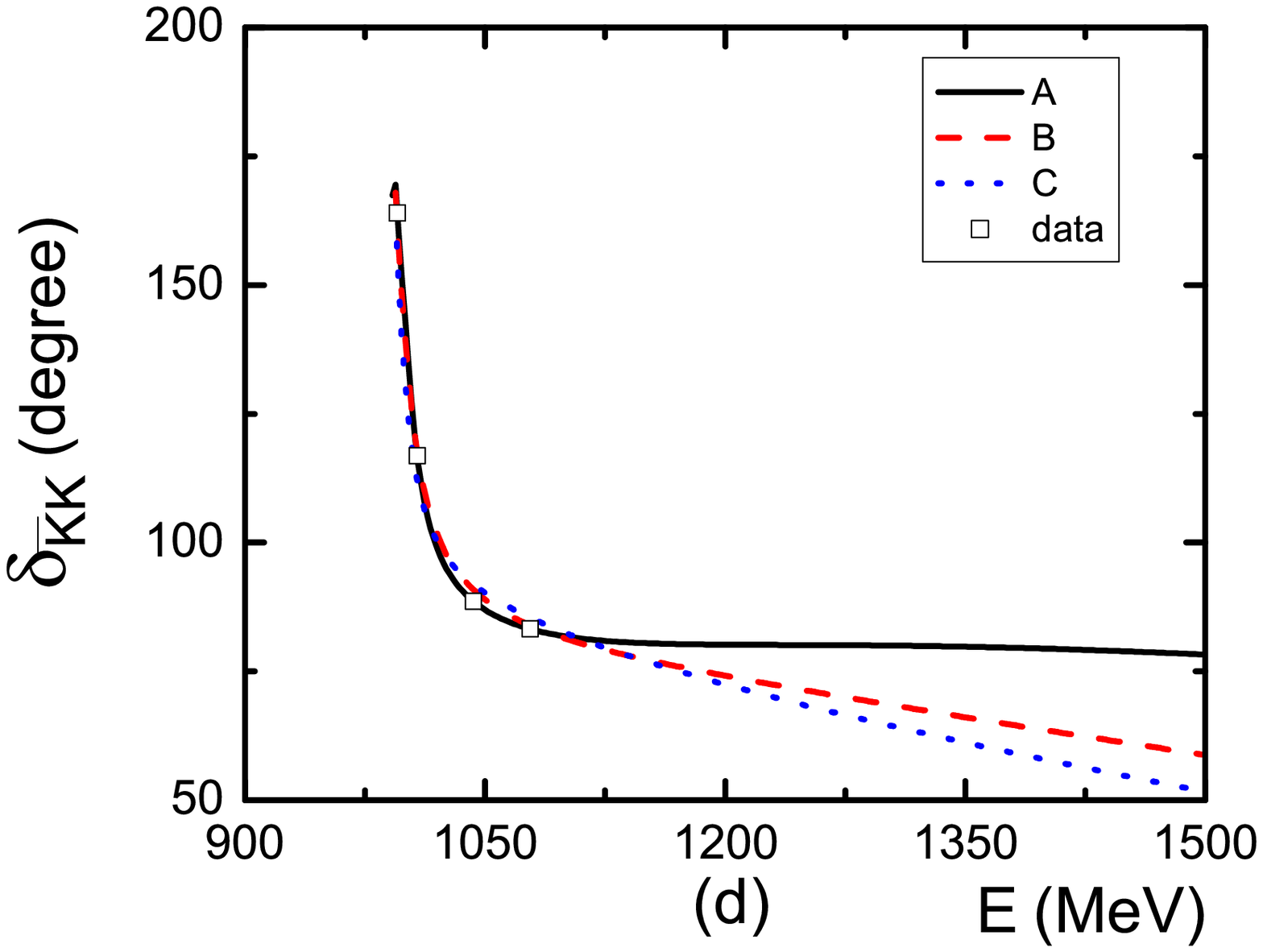}
\caption{(Color online) (a): the spectrum data generated from $1b-2c$ model. (b-d): the
phase shifts and the inelasticity calculated from the two-channels model with
the parametrization $A$ ($1b-2c$ model), $B$ and $C$
specified in Eqs.(\ref{eq:geqa})-(\ref{eq:veqc}) are
compared with the data (from $1b-2c$ model).} \label{fg:fit216}
\end{center}
\end{figure}

\begin{figure}[htbp] \vspace{-0.cm}
\begin{center}
\includegraphics[width=0.49\columnwidth]{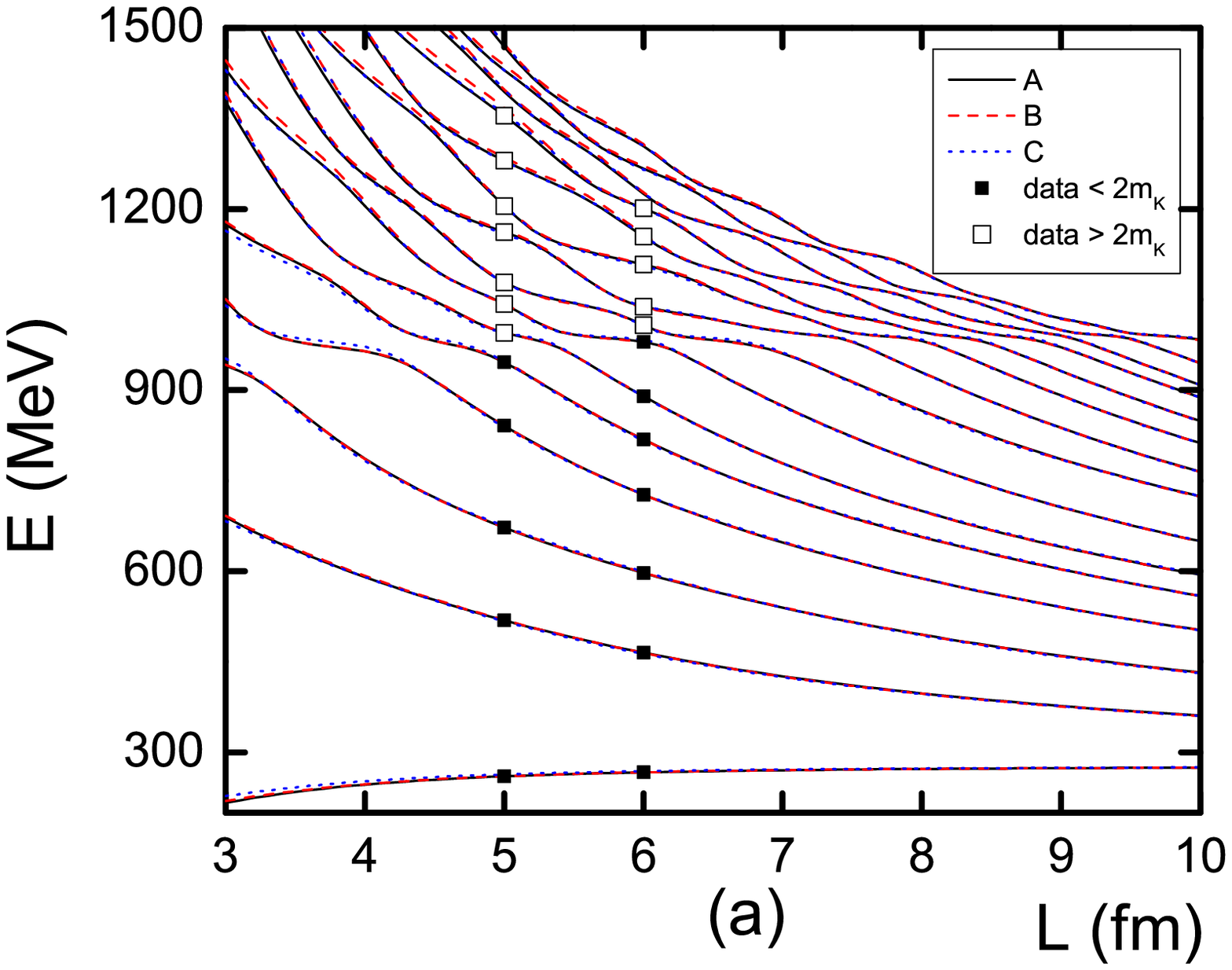}
\includegraphics[width=0.49\columnwidth]{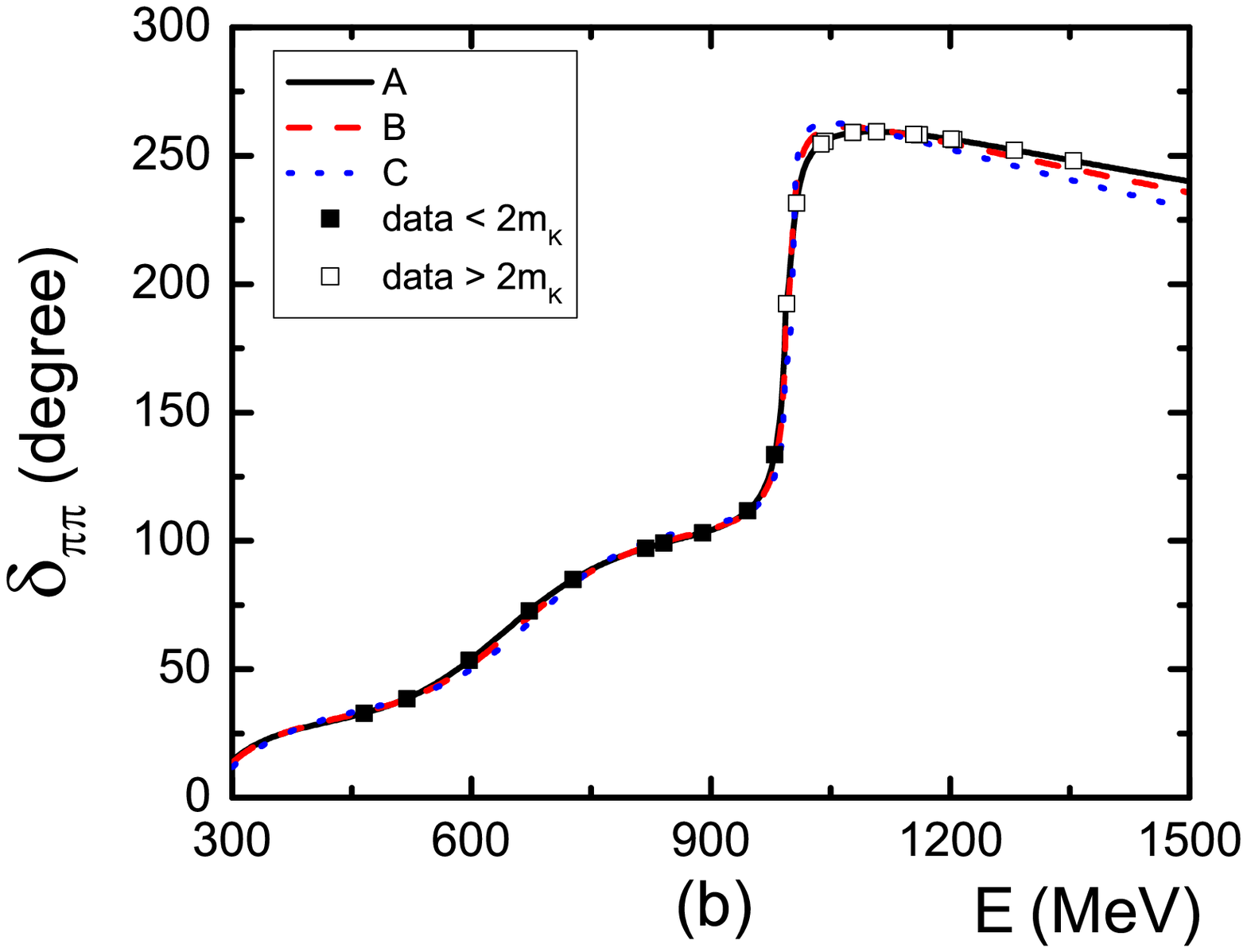}
\includegraphics[width=0.49\columnwidth]{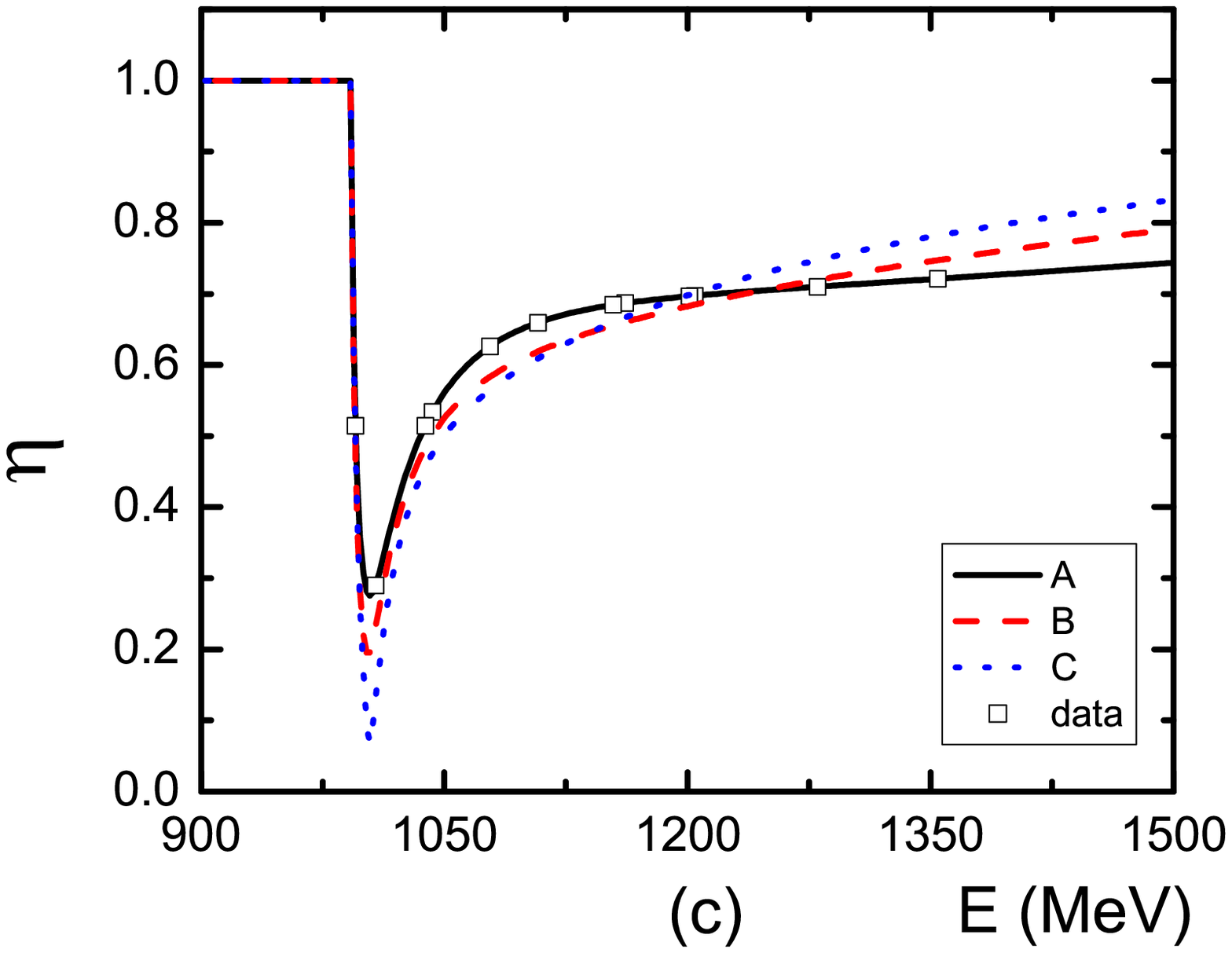}
\includegraphics[width=0.49\columnwidth]{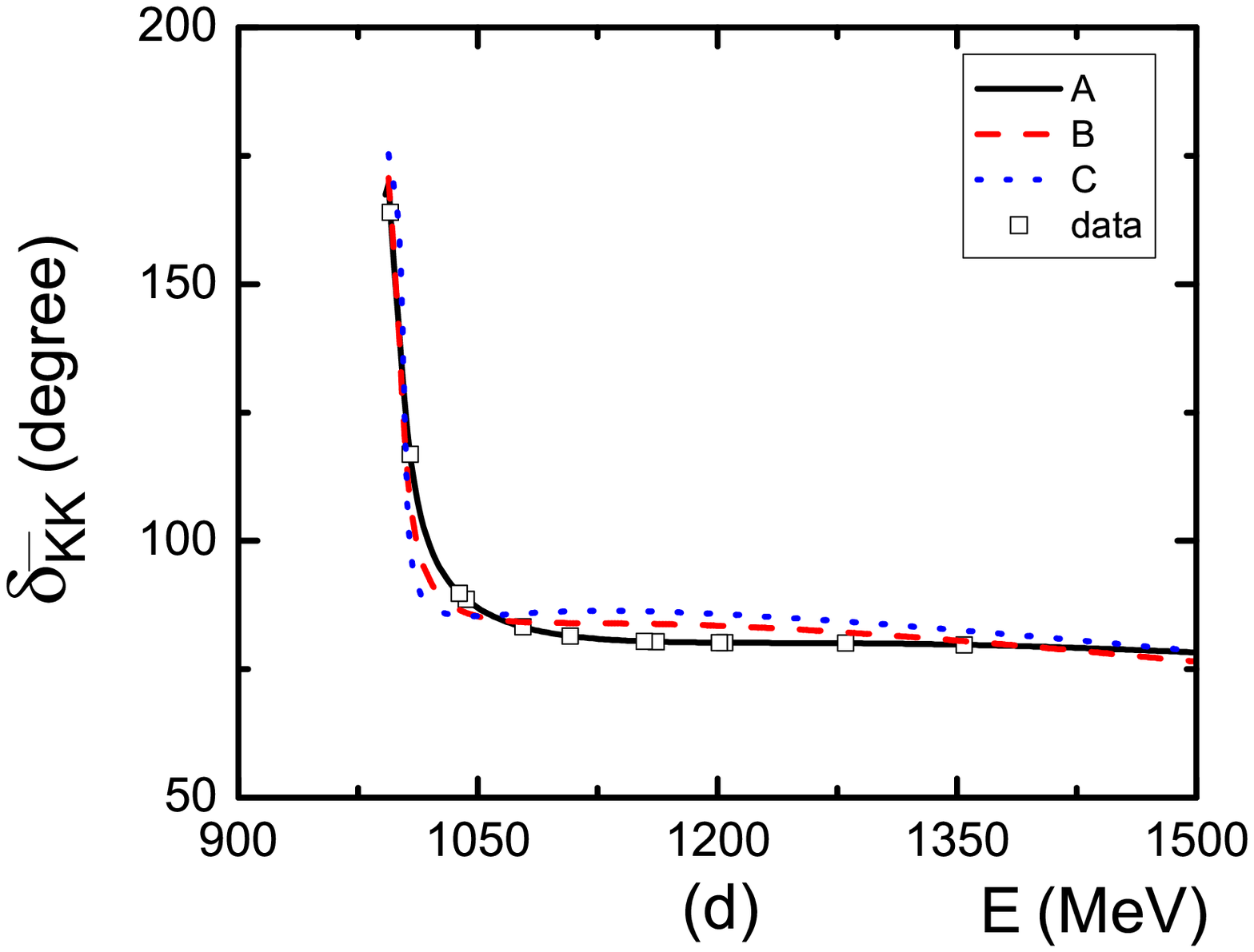}
\caption{(Color online) (a): the spectrum data generated from $1b-2c$ model. (b-d): the
phase shifts and the inelasticity calculated from the one-channels model with
the parametrization $A$ ($1b-2c$ model), $B$ and $C$
specified in Eqs.(\ref{eq:geqa})-(\ref{eq:veqc}) are
compared with the data (from $1b-2c$ model).} \label{fg:fit224}
\end{center}
\end{figure}

Here we explore the finite-volume Hamiltonian method for the
coupled-channels system. We generate 16 and 24 spectrum data points
from the $1b-2c$ model constructed in section IV.A by solving
eigenvalue problem defined by Eqs.(\ref{eq:det-1})-(\ref{eq:vfin-1})
for $L=$ 5, 6 fm.  As shown in the left top panel of
Figs.~\ref{fg:fit216} and \ref{fg:fit224},
these spectrum data can be
fit by a Hamiltonian with
the parametrization $B$ or $C$ of the interaction Hamiltonian
specified in Eqs.~(\ref{eq:geqb})--(\ref{eq:veqc}).
As in the one-channel case, we assign a 1 MeV error for each spectrum
data point in these fits. We see in
Figs.~\ref{fg:fit216}--\ref{fg:fit224} that the phase shifts
$\delta_{\pi\pi}$ and $\delta_{K\bar{K}}$ and inelasticity $\eta$
calculated from the determined Hamiltonians agree well with data (from
model $1b-2c$) in the energy region where the spectrum data are
fitted. Similar to the one-channel case, the predicted phase shifts
deviate from each other outside the energy range of the fitted
spectrum data. We thus conclude that the finite-volume Hamiltonian
offers a method to directly extract the scattering parameters from
numerical simulation.
Furthermore, the method is largely independent of the form of the
Hamiltonian. One should caution that the resulting Hamiltonian can
only be reliably used to predict the scattering observables in the
energy region where the lattice spectra are fit --- as also seen in
the single-channel case.

Here we point out an important difference with the approach using the
two-channel L\"uscher's formula.
As we discussed in section IV, the two-channel L\"uscher formula,
Eq.~(\ref{eq:ev2c}), needs three spectrum data points at the same
energy to calculate two phase shifts and inelasticity. Thus the
spectrum data (open squares) in the left top panel of
Figs. \ref{fg:fit216} and \ref{fg:fit224} are not sufficient to
apply the L\"uscher's method. One thus requires many more
calculations to get a spectrum like the open squares shown in
Fig.\ref{fg:com2} in section IV.
For a given $E$, we
need to get results for three values of $L$, which can be  chosen only after some searches,
since we don't know the
spectrum for each $L$ before the calculation is finished.
Alternatively, the finite-box Hamiltonian method offers a method to
interpolate the lattice spectra with a minimal set of
volumes. Further, the quality of the extraction will naturally improve
more simulation results.

Finally, regarding the relative phase ambiguity mentioned above
\cite{Berkowitz:2012xq}, in the present context of the Hamiltonian
formulation, the finite volume spectra cannot fix the relative sign of the
resonance coupling to different channels, Eq.~(\ref{eq:g2}), nor the
sign of the off-diagonal terms in the direct interaction,
Eq.~(\ref{eq:v2}). Again, these signs only act to constrain the
relative phase between $\delta_\pipi$ and $\delta_\KKbar$ but do
not influence the energy dependence or the isolation of the resonance
pole position.

\section{Spectra from $\pi\pi$ data}
As a final investigation for the present study, we comment on the
possibility of lattice QCD providing the necessary knowledge to
improve on phenomenological scattering parameterisations.

\begin{figure}[htbp] \vspace{-0.cm}
\begin{center}
\includegraphics[width=0.3\columnwidth]{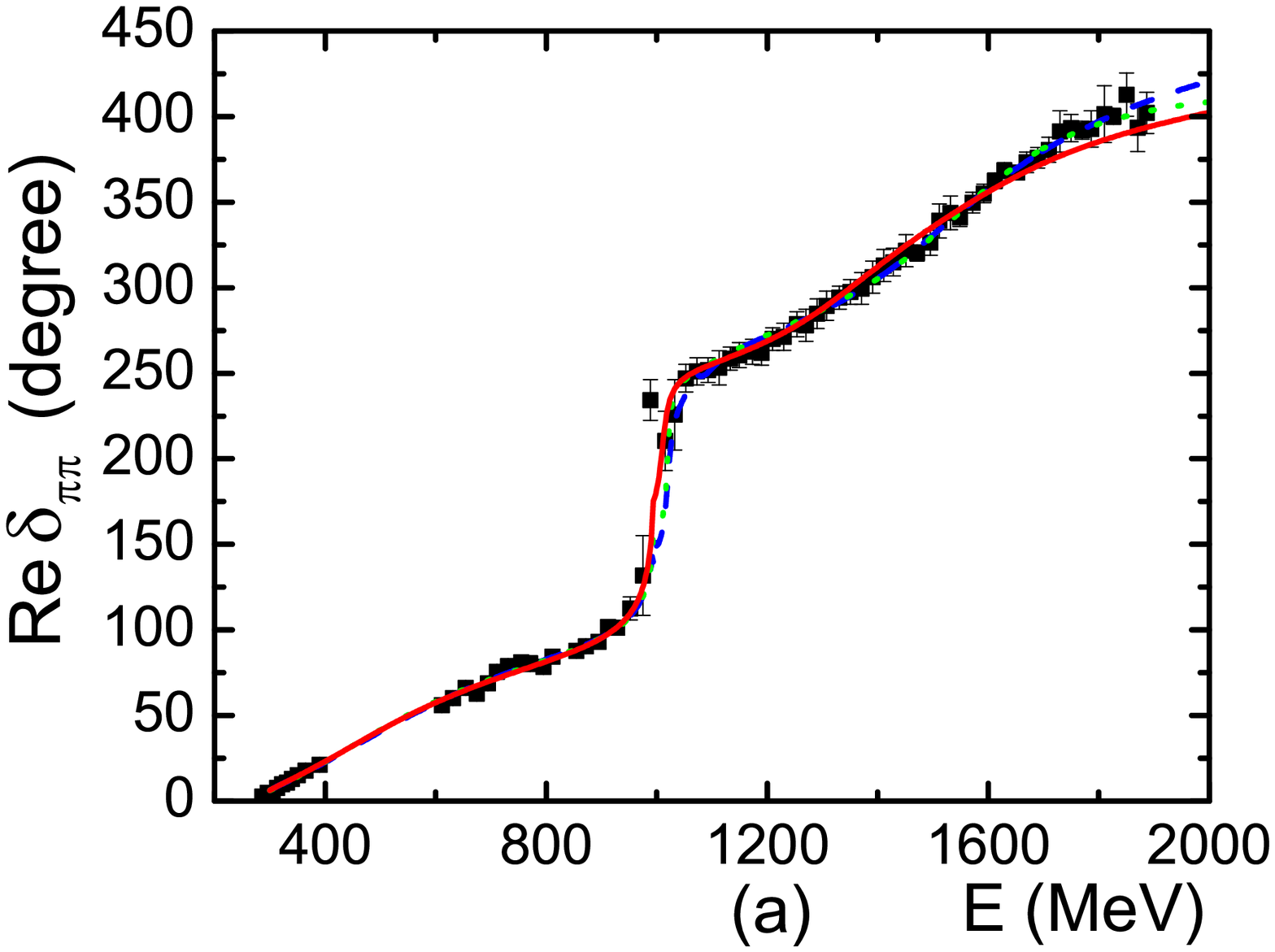}
\includegraphics[width=0.3\columnwidth]{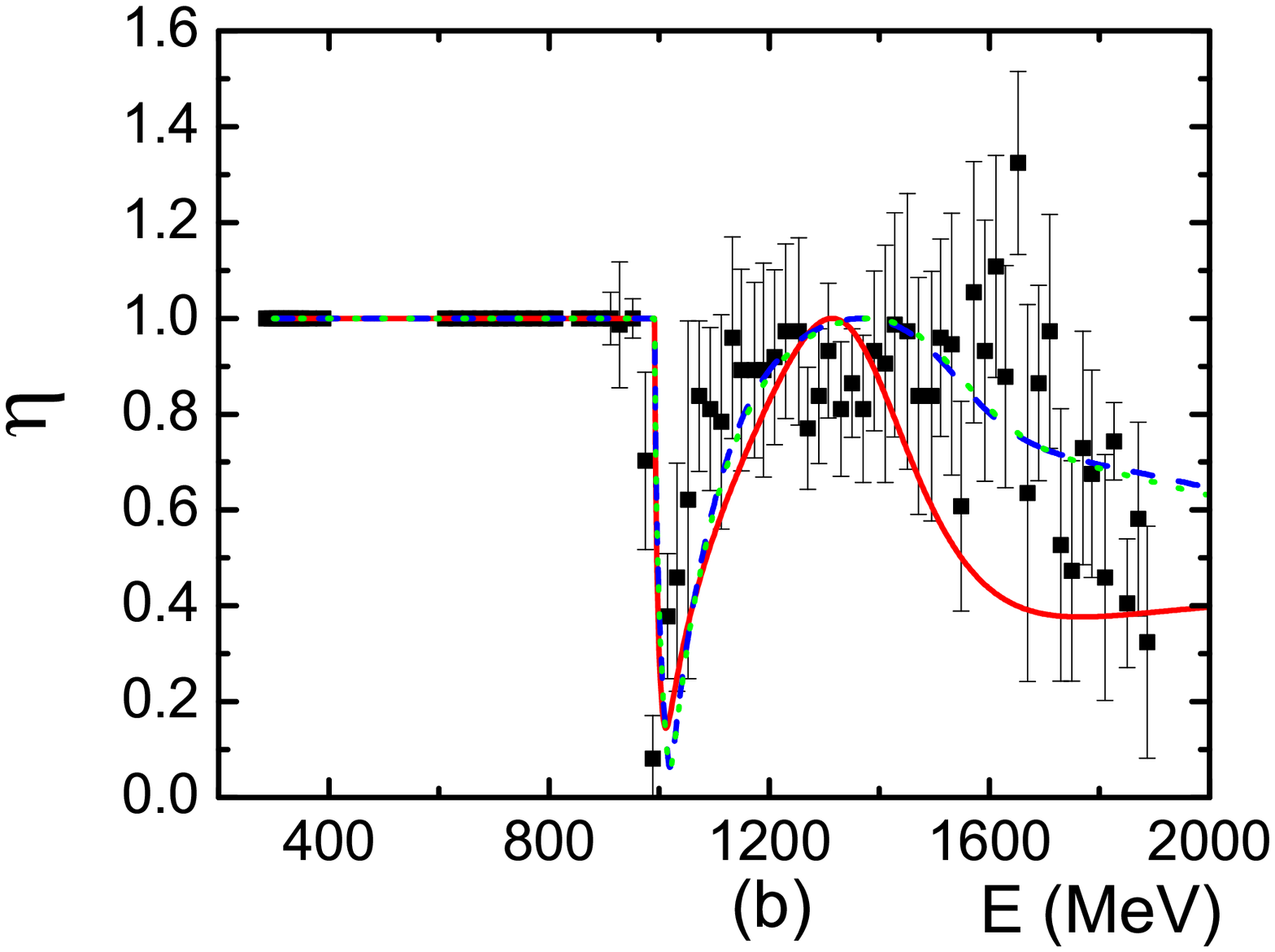}
\includegraphics[width=0.3\columnwidth]{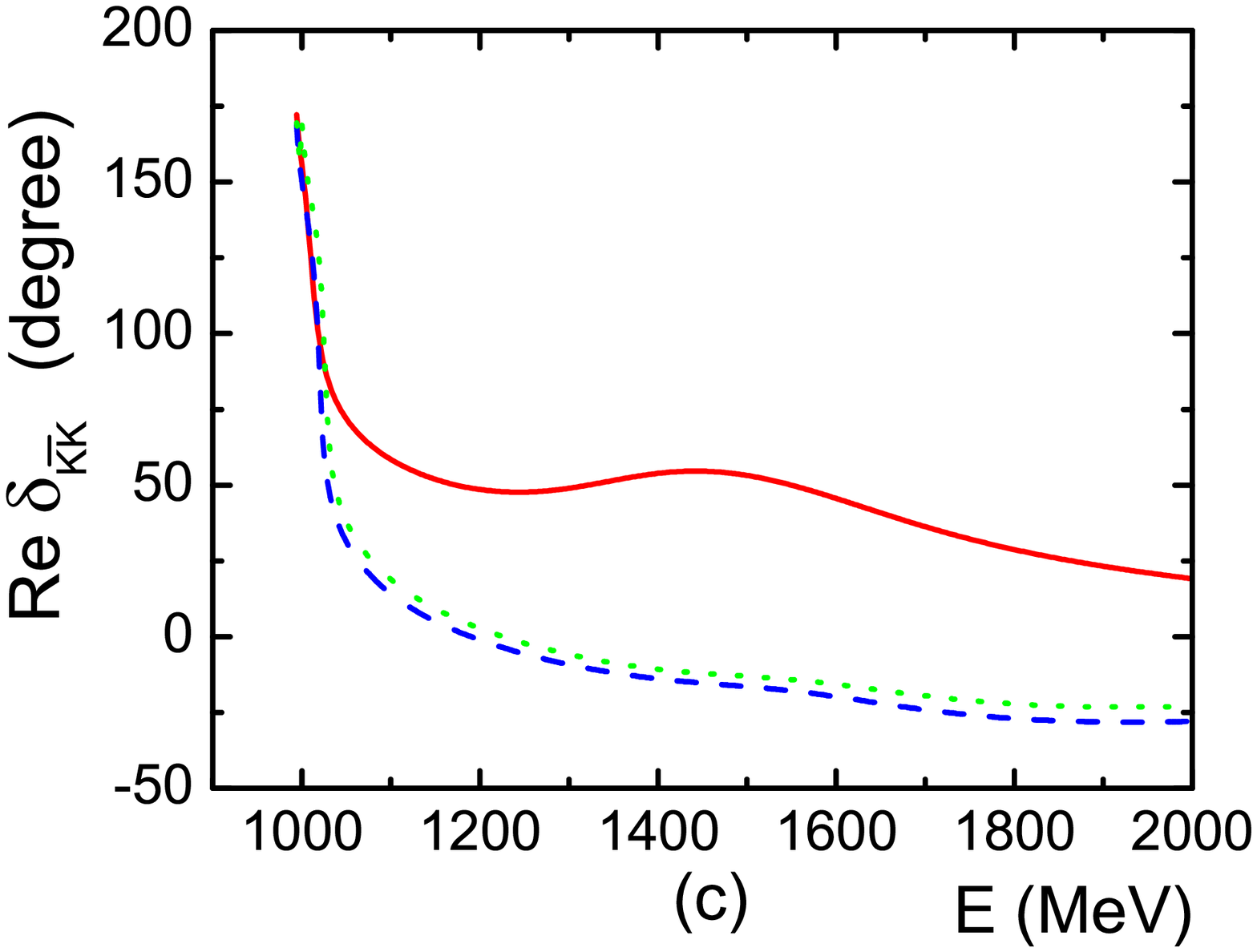}
\caption{(Color online) The phase shifts $\delta_{\pi\pi}$ and $\delta_{K\bar{K}}$,
 and inelasticity $\eta$ of
s-wave $\pi\pi$ scattering in the $J^{IP}=1^{0+}$ partial wave.
The solid squares are the experiment data. The red solid,
blue dashed and green dotted lines are from the NKLS model, Model B, and Model
 C, respectively.} \label{fg:swavefitall}
\end{center}
\end{figure}

\begin{figure}[htbp] \vspace{-0.cm}
\begin{center}
\includegraphics[width=0.3\columnwidth]{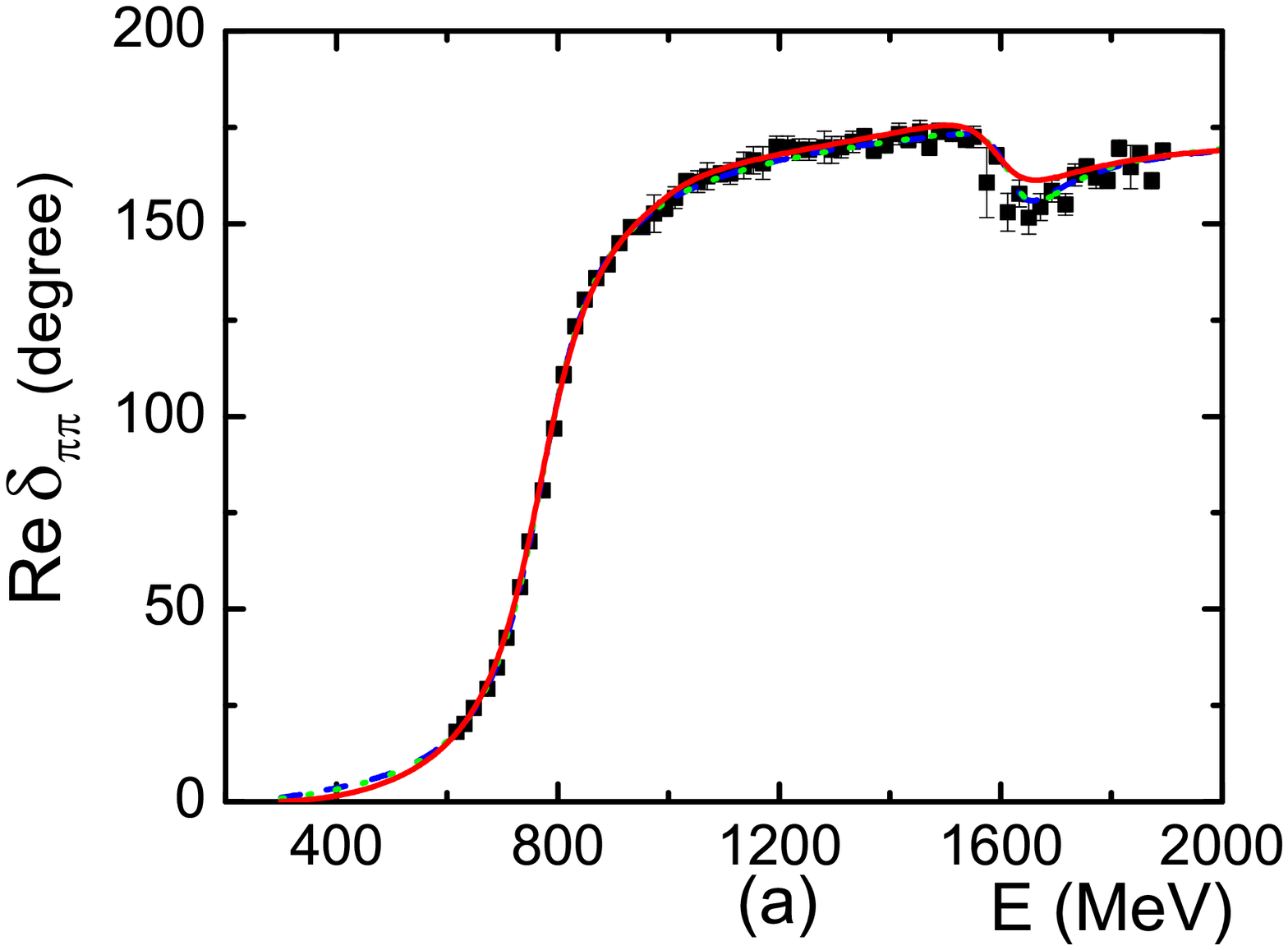}
\includegraphics[width=0.3\columnwidth]{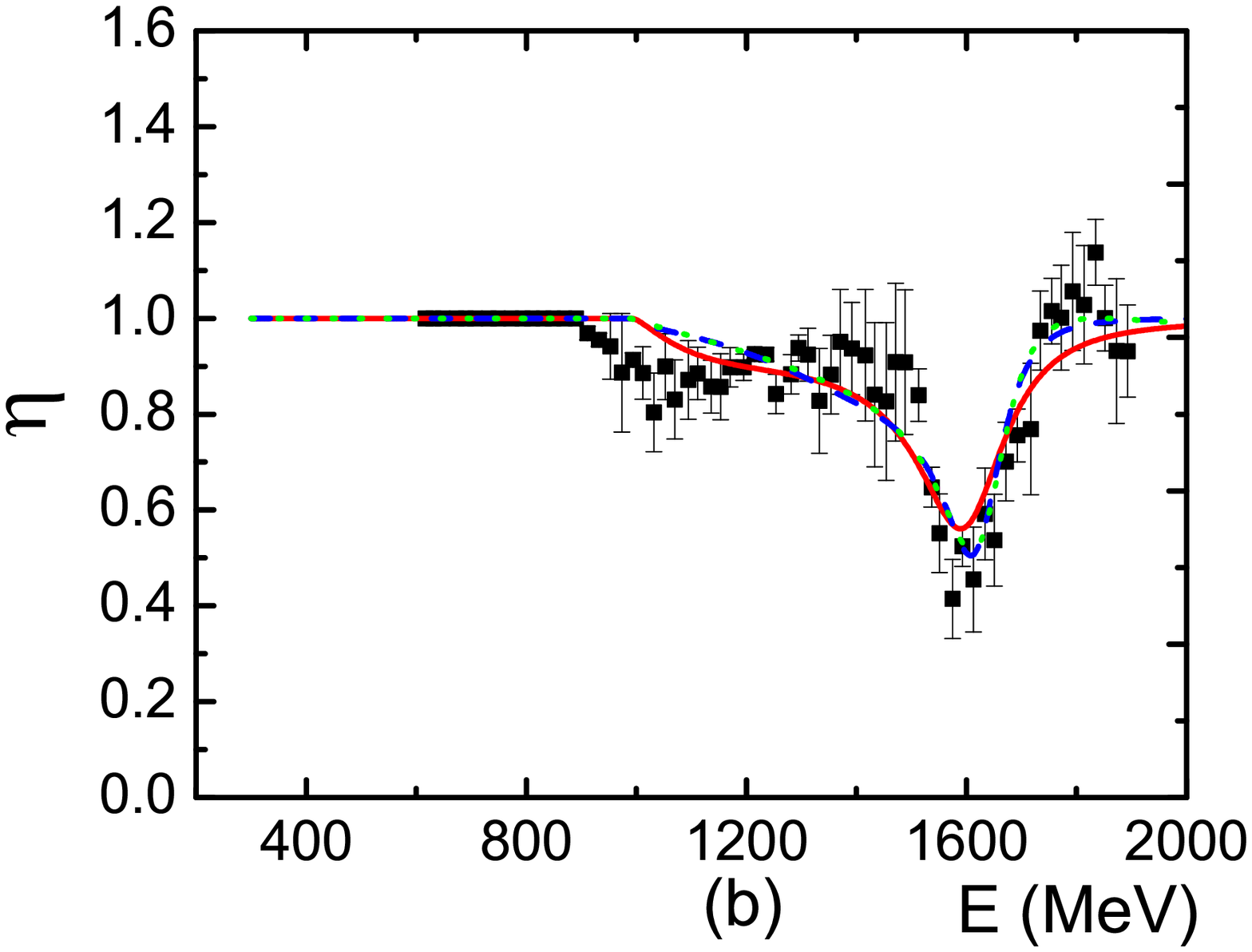}
\includegraphics[width=0.3\columnwidth]{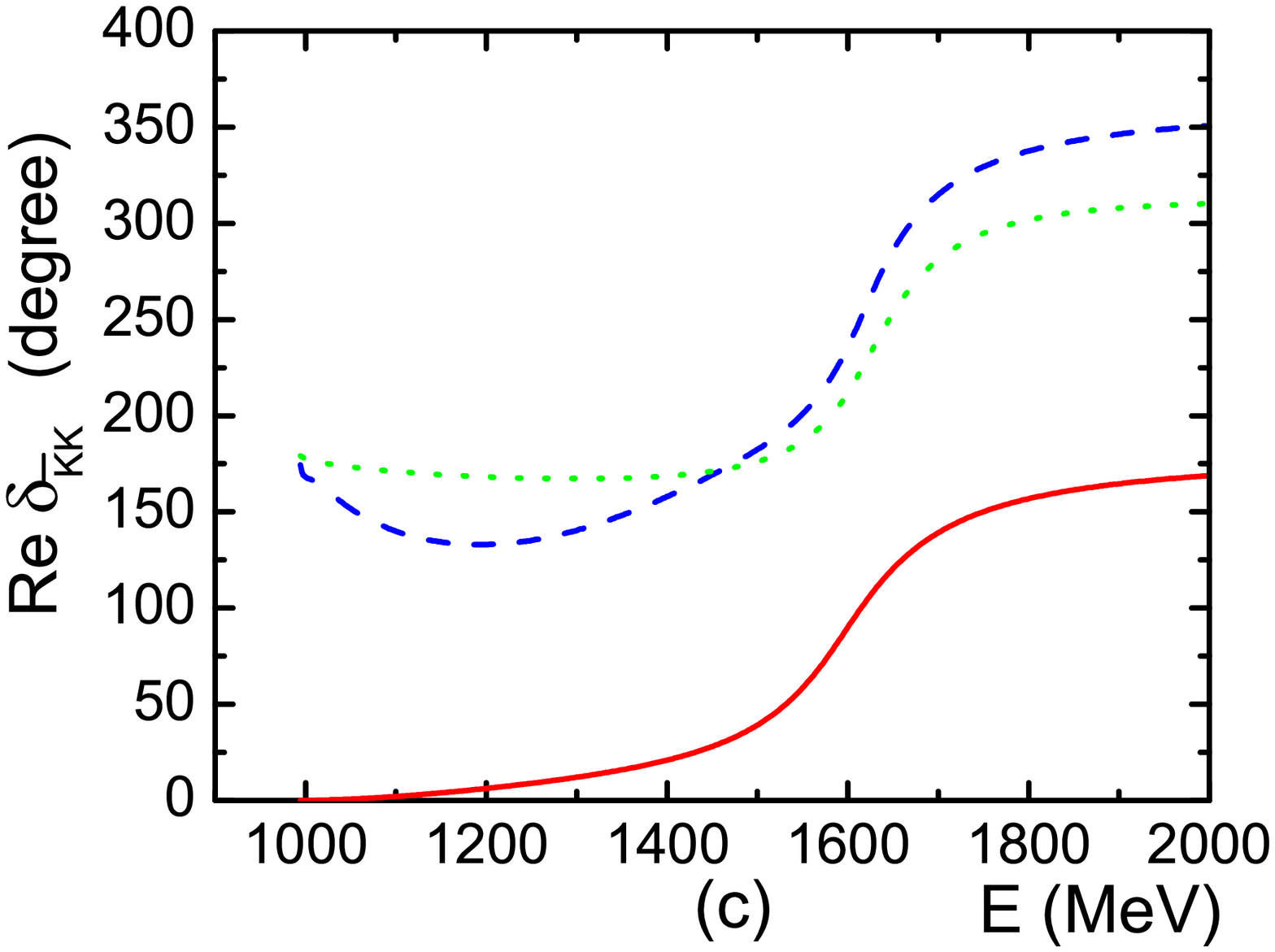}
\caption{(Color online) The phase shifts $\delta_{\pi\pi}$ and $\delta_{K\bar{K}}$,
 and inelasticity $\eta$ of
p-wave$\pi\pi$ scattering in the $J^{IP}=1^{1-}$ partial wave.
The solid squares are the experiment data. The red solid,
blue dashed and green dotted lines are from the NKLS model, Model B, and Model
 C, respectively.} \label{fg:pwavefitall}
\end{center}
\end{figure}

Within the Hamiltonian formulation given in section II, the $\pi\pi$
scattering phase shifts $\delta_{\pi\pi}$ and inelasticity $\eta$ up
to 2 GeV have been fit \cite{Kamano:2011ih} using a model which has
two bare states and includes the $\pi\pi$ and $K\bar{K}$ channels. Its
interaction Hamiltonian only has the vertex interaction $g$ defined in
Eq.~(\ref{eq:int-g}). This model (which we will refer to as NKLS) also
reproduces well the resonance pole positions listed by the Particle Data
Group~\cite{PDG}. We explore a further two models, $B$ and $C$, which further
incorporate the two-body interaction $v$ defined in
Eq.~(\ref{eq:int-v}) with the form Eq.~(\ref{eq:veqa}). These two
solutions give equally good fits to the data of $\delta_{\pi\pi}$ and
inelasticity $\eta$, and the resonance pole positions. The three
models for both S-wave and P-wave scattering are shown in
Figs.~\ref{fg:swavefitall} and \ref{fg:pwavefitall}, with model
parameters listed in Table \ref{tab:para-all}. Note that the
parametrization of the matrix elements of the interactions of NKLS
model are the same as Model $A$ specified in Eq.(\ref{eq:geqa}) and
(\ref{eq:veqa}) except that the parametrization for the p-wave vertex
interaction in the $J^{IP}=1^{1-}$ partial wave is
\begin{eqnarray}
\langle k|g_{\sigma, i}\rangle &=&g_{\sigma, i}(k) \nonumber \\
                      &=&\frac{g_{i}}{\sqrt{m_\pi}}\left(\frac{1}{(1+(c_{\pi\pi}\times k)^2)}\right)^{\frac{3}{2}}\frac{k}{m_\pi},
\end{eqnarray}

\begin{figure}[htbp] \vspace{-0.cm}
\begin{center}
\includegraphics[width=0.49\columnwidth]{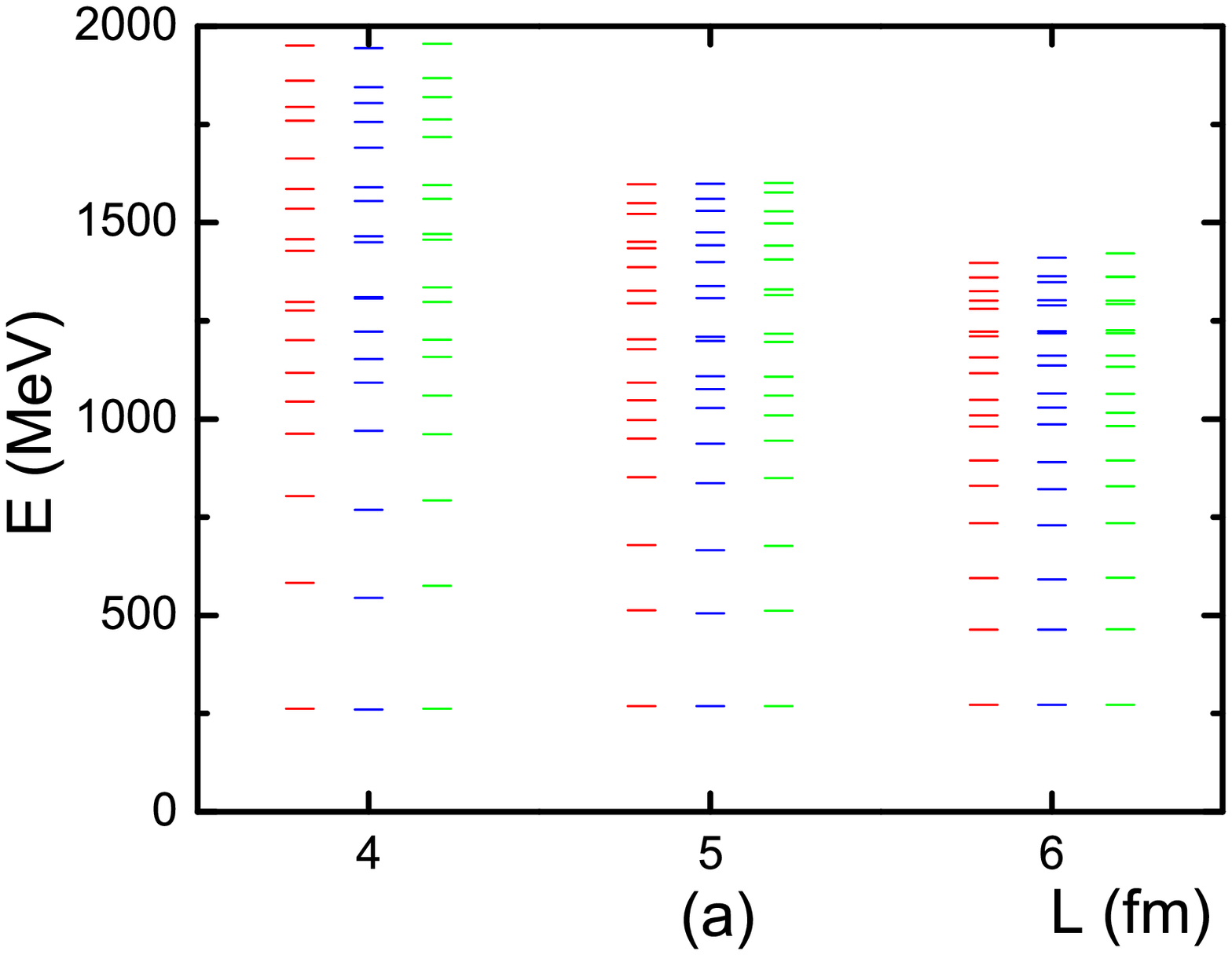}
\includegraphics[width=0.49\columnwidth]{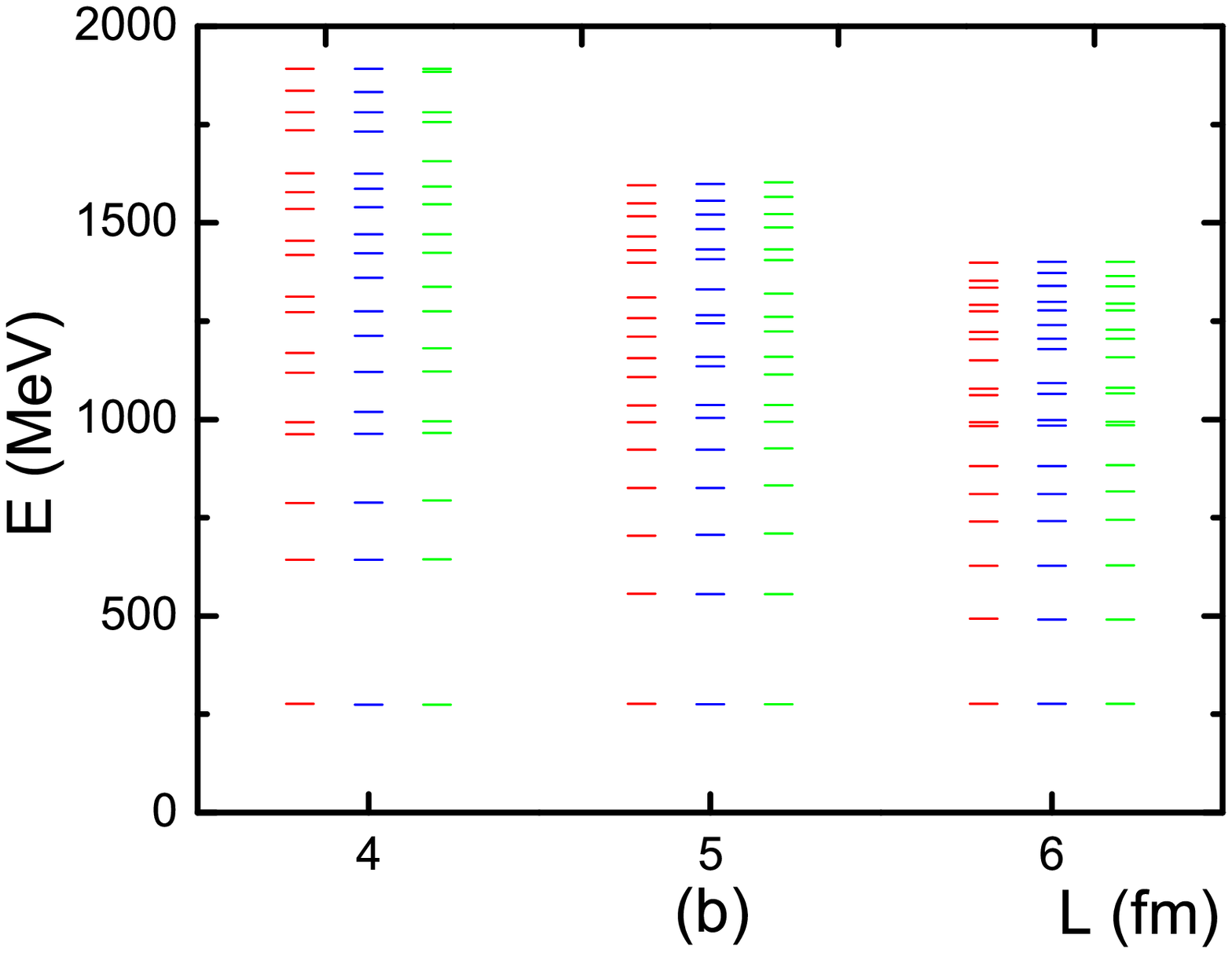}
\caption{(Color online) Spectra for $J^{IP}=0^{0+}$(left) and $J^{IP}=1^{1-}$(right) partial waves
from Models NKLS, Model B and Model C. The spectra have been displaced for clarity.}
\label{fg:spectall}
\end{center}
\end{figure}

As there are no data to constrain the $K\bar{K}$ scattering phase
shifts, this observable displays the largest variation among the model
solutions --- see the right panel of Figs.~\ref{fg:swavefitall} and
\ref{fg:pwavefitall}. We can now explore the sensitivity to this
variation in the predicted finite volume spectra.
These predicted spectra are show in Fig.~\ref{fg:spectall}.
While the spectra are in broad agreement between the models, there
are noticeable differences among the volumes considered.
In particular, on the $4\fm$ box some energy levels see a variation of
up to $50\mev$ between the different model solutions. In
principle, lattice QCD spectra of this order of precision could act to
further constrain this phenomenological model.
One should of course caution that, in principle, there could be
further inelastic channels appearing in the lattice calculation ---
such as 4 pions.

\begin{table}[ht]
     \setlength{\tabcolsep}{0.15cm}
\begin{center}\caption{The parameters of the Hamiltonians from fitting the
phase shift data of $\pi\pi$ scattering in s-wave  $J^{IP}=0^{0+}$ and p-wave $J^{IP}=1^{1-}$ partial waves.}
\begin{tabular}{ccccccc}\hline
                             &   \multicolumn{3}{c}{S-wave}            &   \multicolumn{3}{c}{P-wave} \\
Parameter                    &  NKLS       &  B          &  C          &  NKLS       &  B          &  C       \\
\hline
 $m_{\sigma_1}$(MeV)         & $  1220.0$  & $ 1094.28$  & $ 1300.00$  & $  891.54$  & $ 900.000$  & $ 999.950$  \\
 $g_{\sigma_1\pi\pi}$        & $-0.63474$  & $-0.97085$  & $-0.51274$  & $-0.20583$  & $-0.15561$  & $-0.11669$  \\
 $c_{\sigma_1\pi\pi}$(fm)    & $ 0.44658$  & $ 0.50923$  & $ 0.33070$  & $ 0.49998$  & $ 0.41213$  & $ 0.31296$  \\
 $g_{\sigma_1 K\bar{K}}$     & $ 0.00605$  & $ 1.64234$  & $ 0.07659$  & $ 0.10607$  & $ 0.01010$  & $ 0.00128$  \\
 $c_{\sigma_1 K\bar{K}}$(fm) & $ 0.10012$  & $ 2.29463$  & $ 0.17073$  & $ 0.42241$  & $ 0.17333$  & $ 0.04512$  \\
 $m_{\sigma_2}$(MeV)         & $  2400.0$  & $ 1907.63$  & $ 2318.94$  & $  1840.0$  & $ 1657.66$  & $ 1903.56$  \\
 $g_{\sigma_2\pi\pi}$        & $ 0.49518$  & $ 0.49178$  & $ 1.43296$  & $ 0.01453$  & $ 0.01852$  & $ 0.00517$  \\
 $c_{\sigma_2\pi\pi}$(fm)    & $ 0.20645$  & $ 0.31107$  & $ 0.35299$  & $ 0.10000$  & $ 0.15068$  & $ 0.06607$  \\
 $g_{\sigma_2 K\bar{K}}$     & $-1.17880$  & $-1.53414$  & $-2.50030$  & $ 0.16674$  & $ 2.42851$  & $ 0.10514$  \\
 $c_{\sigma_2 K\bar{K}}$(fm) & $ 0.50000$  & $ 1.06150$  & $ 0.79294$  & $ 0.49993$  & $ 1.71022$  & $ 0.30817$  \\
 $G_{\pi\pi,\;\pi\pi}$       & $ -      $  & $ 0.10000$  & $ 0.10000$  & $ -      $  & $-0.01718$  & $ 0.00024$  \\
 $G_{\pi\pi,\;K\bar{K}}$     & $ -      $  & $-0.00045$  & $-0.07138$  & $ -      $  & $-0.11589$  & $-0.04689$  \\
 $G_{K\bar{K},\;K\bar{K}}$   & $ -      $  & $-0.00016$  & $ 0.09992$  & $ -      $  & $ 0.34790$  & $ 0.02819$  \\
 $d_{\pi\pi}$(fm)            & $ -      $  & $ 0.27088$  & $ 0.18337$  & $ -      $  & $ 0.42441$  & $ 0.26895$  \\
 $d_{K\bar{K}}$(fm)          & $ -      $  & $ 0.00551$  & $ 0.18402$  & $ -      $  & $ 0.41520$  & $ 0.12503$  \\
 $\chi^2$                    & $305     $  & $205     $  & $215     $  & $189     $  & $119     $  & $119     $  \\
 Pole(GeV)                   & $0.43-0.27i$& $0.43-0.32i$& $0.43-0.26i$&$0.77-0.081i$& $0.77-0.075i$& $0.77-0.076i$\\
                             & $1.0-0.010i$& $1.0-0.014i$& $1.0-0.008i$& $1.61-0.11i$& $1.63-0.075i$& $1.65-0.083i$\\
                             & $1.35-0.17i$& $1.51-0.22i$& $1.52-0.20i$& $-$         & $-$          & $-$ \\
\hline\end{tabular}  \label{tab:para-all}
\end{center}
\end{table}

\section{Summary}
We have investigated the finite-volume Hamiltonian method developed in
Ref.~\cite{Hall:2013qba} within several models for $\pi\pi$
scattering. We have demonstrated the equivalence of the finite volume
spectra with the L\"uscher formalism for both a single channel and
also the corresponding generalisation to a coupled-channel system.

We then investigated the practical inversion problem for lattice QCD,
with the aim to determine the physical scattering parameters from the
finite-volume spectra.
%
%
%
The finite-volume Hamiltonian framework offers a robust framework
  for the parameterisation of hadronic interactions to fit lattice
  spectra.
Future work will aim to address outstanding issues, as
  addressed throughout the manuscript, including: the role of angular
  momentum mixing, exponentially suppressed corrections and
  multi-particle inelasticities. The generalisation to moving frames
  will also act to improve the determination of scattering parameters,
  with little additional computational costs.

Based on phenomenological fits to experimental $\pi\pi$ scattering, we
have presented the predicted spectra that one could anticipate seeing
in lattice simulations at the physical pion mass. Here we have
demonstrated that sufficient precision from lattice QCD simulations
would offer the potential to improve the knowledge of these
phenomenological models. This is particularly significant for channels
that are not directly observable in experiment.

Our investigations are based on a rather phenomenological form of the
Hamiltonian.  Thus the constructed Hamiltonian from fitting lattice QCD
spectrum can not be used reliably to predict scattering observables
beyond the energy region where the spectra are fit.
One potential improvement in this general framework would be to
consider more realistic forms of the Hamiltonian, such as those
derived from chiral Lagrangians. This would largely act to improve the
near-threshold behaviour of the interactions, however is beyond the
scope of the present work.

\section{acknowledgment}
%
We wish to thank Ra\'ul Brice\~no for helpful correspondence.
%
This work is supported by the U.S. Department of Energy, Office of
Nuclear Physics Division, under Contract No. DE- AC02-06CH11357. This
research used resources of the National Energy Research Scientific
Computing Center, which is supported by the Office of Science of the
U.S. Department of Energy under Contract No. DE-AC02-05CH11231, and
resources provided on "Fusion", 320-node computing cluster operated by
the Laboratory Computing Resource Center at Argonne National
Laboratory.
This work was also supported by the University of Adelaide and the
Australian Research Council through the ARC Centre of Excellence for
Particle Physics at the Terascale and grants FL0992247 (AWT),
DP140103067, FT120100821 (RDY).

\appendix

\section{L\"uscher summary}
\label{app:lusch}
\subsection{Single channel}
For comparison with L\"uscher's method,
we summarise the formulae relevant to a purely s-wave interaction,
as considered in this manuscript.
It relates each
energy eigenvalues $E$ of the finite box with size $L$ to the
scattering phase shift $\delta$ at energy $E$ by the following
equations:
\begin{eqnarray}
\delta(k)&=&-\phi(q) + n\pi
\label{eq:luch1}
\end{eqnarray}
with the on-shell momenta given by
\begin{equation}
k=\sqrt{E^2/4-m^2_\pi},
\label{eq:onshell}
\end{equation}
and the geometric phase $\phi$ defined by
\begin{equation}
\tan\phi(q)=-\frac{q\pi^{3/2}}{\ZZ_{00}(1;q^2)}\,,
\label{eq:phi}
\end{equation}
expressed in terms of the lattice momenta
\begin{equation}
q=\frac{kL}{2\pi}.
\end{equation}
The generalized zeta function is defined by
\begin{equation}
\ZZ_{00}(1;q^2)=\frac{1}{\sqrt{4\pi}}\sum_{\vec{n}\in \mathbb{Z}^3} (\vec{n}^2-q^2)^{-1}\,,
\label{eq:luch3}
\end{equation}
defined with an appropriate regularisation of the divergent sum (see
eg.~\cite{Luscher:1990ux} for discussion).
Numerically, a convenient representation for the evaluation of the
regularised form is given by
\begin{eqnarray}
\ZZ_{00}(1;q^2)&=&\frac{1}{\sqrt{4\pi}}\left(-\frac{1}{q^2}-8.91363292+16.53231596q^2
+\sum_{\vec{n}\in \mathbb{Z}^3,\vec{n}\neq0}\frac{q^4}{\vec{n}^4(\vec{n}^2-q^2)}\right).
\label{eq:luch4}
\end{eqnarray}

\subsection{Coupled channel}
\label{app:multi}
At energies above the $K \bar{K}$ threshold, we need L\"uscher's
method for two open channels, as developed in Ref.\cite{He:2005ey}.
For the considered $\pi\pi$ and $K\bar{K}$ channels, the $S$-matrix is
defined by
\begin{eqnarray}
S&=&\left( \begin{array}{cc}
\eta e^{2i\delta_{\pi\pi}}                                 & i\sqrt{1-\eta^2} e^{i(\delta_{\pi\pi}+\delta_{K\bar{K}})}                            \\
i\sqrt{1-\eta^2} e^{i(\delta_{\pi\pi}+\delta_{K\bar{K}})}  &\eta e^{2i\delta_{K\bar{K}}}
\end{array} \right), \label{eq:luch1a}
\end{eqnarray}
where the phase shifts $\delta_{\pi\pi}$ and $\delta_{K\bar{K}}$ and inelasticity
$\eta$ at each $E$ are related to the box size $L$ by the following relation
%
%
\begin{align}
&\cos\left[\phi(q_{\pi\pi})+\phi(q_\KKbar)-\delta_{\pi\pi}(E)-\delta_{K\bar{K}}(E)\right]\non\\
&\quad-\eta(E)\cos\left[\phi(q_{\pi\pi})-\phi(q_\KKbar)-\delta_{\pi\pi}(E)+\delta_{K\bar{K}}(E)\right]=0.
\label{eq:luch2a}
\end{align}
where $\phi(q_{\alpha})$ is defined as Eq.~(\ref{eq:phi}), and
\begin{eqnarray}
q_{\alpha}&=&\frac{k_\alpha(E)L}{2\pi}. \label{eq:luch4a}
\end{eqnarray}

\section{Relationship between the Hamiltonian and L\"uscher quantisations}
\label{app:relationship}
The L\"uscher formalism has established that the finite volume
spectrum of multi-particle states is determined by an eigenvalue
equation involving just the $S$-matrix of the corresponding theory ---
up to corrections which are exponentially suppressed in $m L$ for
large volumes. This has been derived on the basis of the underlying
fields satisfy the periodicity of the lattice and that the
interactions are finite-range in nature, limited by a mass scale $m$
(typically the lightest particle degree of freedom present in the
system). The Hamiltonian formulation presented here, and previously in
Ref.~\cite{Hall:2013qba}, has an interaction which is finite ranged
and the fields themselves are quantised to satisfy the lattice
periodicity. Therefore, in terms of the quantisation condition on the
spectra, the Hamiltonian no more than an explicit realisation of the
general conditions considered by L\"uscher.

In the Sec.~\ref{sIIIc} and ~\ref{sIVc}, we have numerically
demonstrated the correspondence between the Hamiltonian and L\"uscher
spectra.  In this appendix, for the case of a simple idealised system
we provide an analytic derivation of the connection between the
L\"uscher and Hamiltonian formalisms.


\subsection{Hamiltonian quantisation}
From the Eqs.~(\ref{eq:h01},\ref{eq:hi1},\ref{eq:gfin}), the Hamiltonian
matrix for the single-channel case with $v=0$ is given by:
\begin{eqnarray}
[H]_{N+1}&=&\left( \begin{array}{cccccccc}
m_\sigma                 & g^{fin}_{\pi\pi}(k_0)               & g^{fin}_{\pi\pi}(k_1)          & \cdots \\
g^{fin}_{\pi\pi}(k_0)   & 2\sqrt{k^2_0+m^2_{\pi}}             & 0                              & \cdots \\
g^{fin}_{\pi\pi}(k_1)   & 0                                   & 2\sqrt{k^2_1+m^2_{\pi}}        & \cdots \\
\vdots                       & \vdots                                   & \vdots           & \ddots
\end{array} \right).
\end{eqnarray}
The eigenvalue $E_i$ of above matrix is satisfied following equation:
\begin{eqnarray}
E_i-m_\sigma &=& \left(\frac{2\pi}{L}\right)^3\frac{1}{4\pi} \sum_{\vec{k}_n = \frac{2\pi}{L}\vec{n},\,\vec{n}\in \mathbb{Z}^3}
\frac{g^2_{\sigma,\pi\pi}(k_n)}{E_i-2E_{\pi}(k_n)}.\label{eq:Ham}
\end{eqnarray}
This can be rearranged to the form
\begin{equation}
E_i-m_\sigma = \left(\frac{2\pi}{L}\right)^3\frac{1}{4\pi} \sum_{\vec{k}_n = \frac{2\pi}{L}\vec{n},\,\vec{n}\in \mathbb{Z}^3}
\left[\frac{E_i g^2_{\sigma,\pi\pi}(k_n)}{2(k_i^2-k_n^2)}-\frac{g^2_{\sigma,\pi\pi}(k_n)}{E_i+2E_\pi(k_n)}\right],
\end{equation}
with $k_i$ implicitly defined by $E_i=2\sqrt{m_\pi^2+k_i^2}$. To
highlight the comparison with the L\"uscher eigenvalue equation, we
further isolate the pole term,
\begin{equation}
E_i-m_\sigma = \left(\frac{2\pi}{L}\right)^3\frac{1}{4\pi} \sum_{\vec{k}_n = \frac{2\pi}{L}\vec{n},\,\vec{n}\in \mathbb{Z}^3}
\left[\frac{E_i g^2_{\sigma,\pi\pi}(k_i)}{2(k_i^2-\vec{k}_n^2)}+\frac{E_i \left(g^2_{\sigma,\pi\pi}(k_n)-g^2_{\sigma,\pi\pi}(k_i)\right)}{2(k_i^2-\vec{k}_n^2)}
-\frac{g^2_{\sigma,\pi\pi}(k_n)}{E_i+2E_\pi(k_n)}\right].
\end{equation}
The last two terms of the RHS have no singularities, and
hence this discrete sum can be approximated by the continuum intergal
(up to corrections of the order of $e^{-m_\pi L}$).
Moving the principal value parts of the sum to the LHS
\begin{equation}
E_i-m_\sigma -\Sigma_L^{\rm PV}(E_i) = \frac{E_ig^2_{\sigma,\pi\pi}(k_i)}{8\pi} \left(\frac{2\pi}{L}\right)^3
\sum_{\vec{k}_n = \frac{2\pi}{L}\vec{n},\,\vec{n}\in \mathbb{Z}^3}
\frac{1}{(k_i^2-\vec{k}_n^2)},\label{eq:Ham2}
\end{equation}
where $\Sigma_L^{\rm PV}$ denotes the finite-volume implementation of
the real part of the self energy. We do note that in performing this
separation we have introduced ultraviolet divergences to both sides of
the equation, these of course exactly cancel each other and have no
significance in determining the infrared properties associated with
the finite volume quantisation.

\subsection{L\"uscher quantisation}
With the conventional parameterisation of the S matrix, $S=\exp(2i\delta)$,
%
and our definition of the T=matix given by
Eqs.~(\ref{eq:ST1}-\ref{eq:ST3}), the phase shift $\delta$ can be
directly evaluted from the equation
\begin{eqnarray}
k_{on}\cot\delta(E)&=& -\frac{4}{\pi E}t^{-1}(E)+ik_{on},\label{eq:deltat}
\end{eqnarray}
where $k_{on}$ is the on-shell momentum of single pion for total
center mass energy E.

With $H_I = g$, the $t$ of $\pi\pi$ channel is:
\begin{align}
t(E)&= \frac{g^2_{\sigma,\pi\pi}}{E-m_\sigma-\Sigma(E)},\\
\Sigma(E)&= \int k^2dk
\frac{g^2_{\sigma,\pi\pi}(k)}{E-2E_{\pi}(k)+i\varepsilon},
\end{align}
and hence the phase shift is given by:
\begin{eqnarray}
k_{on}\cot\delta(E)&=& \frac{-4}{\pi E}\frac{1}{g^2_{\sigma,\pi\pi}(k_{on})}\left(E-m_\sigma-\Sigma^{\rm PV}(E)]\right).\label{eq:phasesigma}
\end{eqnarray}

Neglecting the influence of the partial wave mixing, and any
exponentially suppressed corrections, the eigenvalue equation of the
L\"uscher formalism can be expressed as
\begin{equation}
k_{on}\cot\delta(k_{on})=\frac{2}{\sqrt{\pi}L}\ZZ_{00}(1;q_{on}^2),\label{eq:phaselusch}
\end{equation}
with $q_{on}=k_{on}L/(2\pi)$. Equating Eq.~(\ref{eq:phaselusch}) with
the exact model phase shift of Eq.~(\ref{eq:phasesigma}) with some
straightforward manipulation yields:
%
%
\begin{align}
E-m_\sigma-\Sigma^{\rm PV}(E)=\frac{E_ig^2_{\sigma,\pi\pi}(k_i)}{8\pi} \left(\frac{2\pi}{L}\right)^3\sum_{\vec{k}_n = \frac{2\pi}{L}\vec{n},\,\vec{n}\in \mathbb{Z}^3}\frac{1}{(k_i^2-\vec{k}^2)}.
\end{align}
This we recognise as the same eigenvalue equation described by the
Hamiltonian formulation in Eq.~(\ref{eq:Ham2}), up to the difference
$\Sigma^{\rm PV}-\Sigma_L^{\rm PV}$ --- which is known to be
exponentially suppressed.

\end{document}